\documentclass[pdflatex,sn-nature]{sn-jnl}% Style for submissions to Nature Portfolio journals
\usepackage{multibib}
\newcites{sup}{Supplementary References}
\usepackage{graphicx}%
\usepackage{multirow}%
\usepackage{amsmath,amssymb,amsfonts}%
\usepackage{amsthm}%
\usepackage{mathrsfs}%
\usepackage[title]{appendix}%
\usepackage{xcolor}%
\usepackage{textcomp}%
\usepackage{manyfoot}%
\usepackage{booktabs}%
\usepackage{algorithm}%
\usepackage{algorithmicx}%
\usepackage{algpseudocode}%
\usepackage{listings}%
\usepackage{tabularx} % Load the package
\usepackage{color}
\usepackage{tabularray}
\usepackage{float}
\usepackage{xr}
\usepackage{enumitem}
\usepackage{booktabs}

\theoremstyle{thmstyleone}%
\theoremstyle{thmstyletwo}%

\theoremstyle{thmstylethree}%

\begin{document}

%\title[Article Title]{Complexity and Homophily in Olfactory Cultural Evolution of Perfumes}
\def\scititle{
	Cultural Evolution of Perfumes since 1900
}
%A century of Perfumes: Insights from crowd-sourced Olfactory data
% \title[Article Title]{Cultural Evolution in Perfumes since 1900}%\\\textbf{Title 2:} Individual and Social Factors in the Evolution of Perfumes}
\title{\scititle}

\author*[1]{\fnm{Vahid} \sur{Satarifard}}\email{vahid.satarifard@yale.edu}
\author[2]{\fnm{Fabian} \sur{Baumann}}
\author[3,4]{\fnm{Geetanjali} \sur{Minsky}}
\author[5]{\fnm{Laura} \sur{Sisson}}
\author[6]{\fnm{Lou M.} \sur{Haux}}
\author[7,8]{\fnm{Christophe} \sur{Laudamiel}} 
\author[1]{\fnm{Nicholas A.} \sur{Christakis}} 

\affil[1]{\orgdiv{Human Nature Lab}, \orgname{Yale University}, \city{New Haven}, \state{CT}, \country{USA}}
\affil[2]{\orgdiv{Department of Biology}, \orgname{University of Pennsylvania}, \city{Philadelphia}, \state{PA}, \country{USA}}
\affil[3]{\orgdiv{Tangible Media Group}, \orgname{MIT Media Lab}, \city{Cambridge}, \state{MA}, \country{USA}}
\affil[4]{\orgdiv{Institute for Future Technologies}, \orgname{De Vinci Higher Education}, \city{Paris}, \country{France}}
\affil[5]{\orgname{Patina}, \city{New York}, \state{NY}, \country{USA}}
\affil[6]{\orgdiv{Center for Adaptive Rationality}, \orgname{Max Planck Institute for Human Development}, \city{Berlin}, \country{Germany}}
\affil[7]{\orgdiv{DreamAir LLC}, \orgname{Department of Scent Engineering}, \city{New York}, \state{NY}, \country{USA}}
\affil[8]{\orgdiv{Osmo Labs, PBC}, \city{New Jersey}, \state{NJ}, \country{USA}}

\maketitle

% \section{Main}\label{sec1}

\section*{Abstract}
Perfumes are cultural artifacts and works of sensory art, composed from a finite, recombinable palette of notes that together evoke a distinctive scent impression. Here, we assemble the largest perfume corpus compiled to date, spanning multiple independent databases from 1900 to 2024, and study its evolution through a multidisciplinary computational framework. We first characterize perfumes by properties such as complexity and novelty using crowd-sourced data, finding that compositions have grown more minimalist in their scent profiles, yet more novel in their note combinations, a transition that began in the 1990s. Next, we construct copy-lineage networks and examine how notes are selected across successive time windows. We show that although imitation is pervasive, a growing share of notes drifts free of selection, and original creations retain a measurable quality premium, where they last longer, project further, and earn higher regard. Finally, we construct the collaboration network of master perfumers and show that a perfumer's creative style behaves as a social contagion, transmitted through collaboration and decaying with social distance. We discuss the principal forces shaping this evolution, including regulatory restrictions, cultural change, and the consolidation in the industry. Our findings position perfume as a culturally evolving system, akin to music and fashion, through which societies express and communicate hedonic sensory experiences.
\section{Introduction}

Once considered weak in humans \cite{mcgann2017poor}, olfaction is now recognized as a powerful sensory modality. Humans can discriminate an enormous number of olfactory stimuli \cite{bushdid2014humans}, even though the precise figure remains debated \cite{meister2015dimensionality}. Humans use olfactory cues for intraspecific communication, with important social consequences \cite{stoddart1990scented}. Odor pleasantness has been considered one of the primary perceptual dimensions of olfaction \cite{yeshurun2010odor}, with vital evolutionary implications ranging from food sourcing to mating \cite{stevenson2010initial,wedekind1995mhc}. The perception of odor pleasantness has both universal aspects and culture-specific dimensions \cite{arshamian_perception_2022,sorokowski_is_2024}, and there is growing evidence that these preferences are shaped by both molecular structure \cite{khan_predicting_2007} and socio-cultural context \cite{majid_human_2021}.

Perfumes have served as means of social signaling, medicine, ritual, and creative beauty expression, reflecting some of the aesthetic values of their time and place \cite{reinarz2014past}. The deliberate use of fragrant plant materials may even reach deep into prehistory. For instance, pollen clusters at a Neanderthal burial site have been interpreted as a funerary use of flowers \cite{leroi1975flowers}. Scent was widespread across ancient civilizations, from some of the earliest recorded perfume formulas, attributed to the Mesopotamian perfumer Tapputi-Belatekallim \cite{levey1956babylonian}, to the pleasant aromas of Egyptian mummies \cite{paolin2025ancient}.

Fragrances can potentially intervene in social communication by modulating perceptions of socially relevant cues \cite{allen2019effects}. From the perspective of dual inheritance theory, or gene-culture coevolution \cite{boyd1988culture}, perfume use could thus represent an intentional modification of our chemical signatures, shaping social relationships in a specific cultural manner \cite{havlivcek2012perfume}. Far from being a trivial commodity, when seen across a range of time scales, scent serves as a cultural artifact and a medium of communication that has evolved over time. Yet, unlike other cultural artifacts such as music and language, the cultural evolution of perfumes has not yet received significant attention.

Digitized archives of cultural production allow researchers to study long-term cultural change with breadth and granularity. Across domains such as music \cite{mauch2015evolution,gonzalez2023quantifying}, digitized books \cite{michel2011quantitative}, baby-naming practices \cite{newberry_measuring_2022}, patent innovation \cite{youn_invention_2015}, visual arts \cite{sigaki2018history,lee2020dissecting}, beauty standards \cite{boucherie2026cultural}, and speech acts during the French Revolution \cite{barron2018individuals}, cultural evolution has been computationally traced and modeled \cite{creanza2017cultural,mesoudi2011cultural}. Perfume, however, remains comparatively understudied in this context. Attempts to reconstruct ancient scents, whether from archaeological residues or mummified remains, have already shown how decoding ancient olfactory information can illuminate the evolution of human culture from a new angle \cite{huber2022use,paolin2025ancient}. In addition, industrial fragrance chemistry itself has a rich history of technological transformation \cite{david2023industrial}, leaving environmental traces detectable even in a Caucasus ice core \cite{vecchiato2020great}.

Here, we assemble and harmonize the largest perfume corpus studied to date, comprising approximately 169,000 perfumes compiled from three online databases, with a primary temporal window spanning 1900-2024. We pursue four lines of analysis. First, we characterize the macroscale properties of the corpus, including time series of new perfume releases, the most prominent perfumers and brands, a global map of perfume release, and century-long shifts in the frequency of notes, accords, and gender associations. Second, we describe perfumes by the Shannon entropy of their composition (complexity) and by the co-occurrence surprisal of their notes relative to all prior perfumes (novelty), and we relate these measures to additional perfume attributes derived from crowd-sourced data. 
Third, from resemblance annotations, we build time-directed copy-lineage networks that treat the perfumes as an evolving cultural population, and we use the Price equation to quantify cultural selection on individual notes, defining a perfume's fitness as the number of later perfumes that imitate it. Finally, we construct the collaboration network of master perfumers and examine how social contagion shapes the evolution of perfumery style. These analyses establish perfume as a high-resolution model system for the quantitative study of sensory culture and its evolution.

\section{Results}\label{sec2}
\subsection{Overview of the Perfume Dataset}

\begin{figure}[!htbp]
	\centering
	\includegraphics[width=1\textwidth]{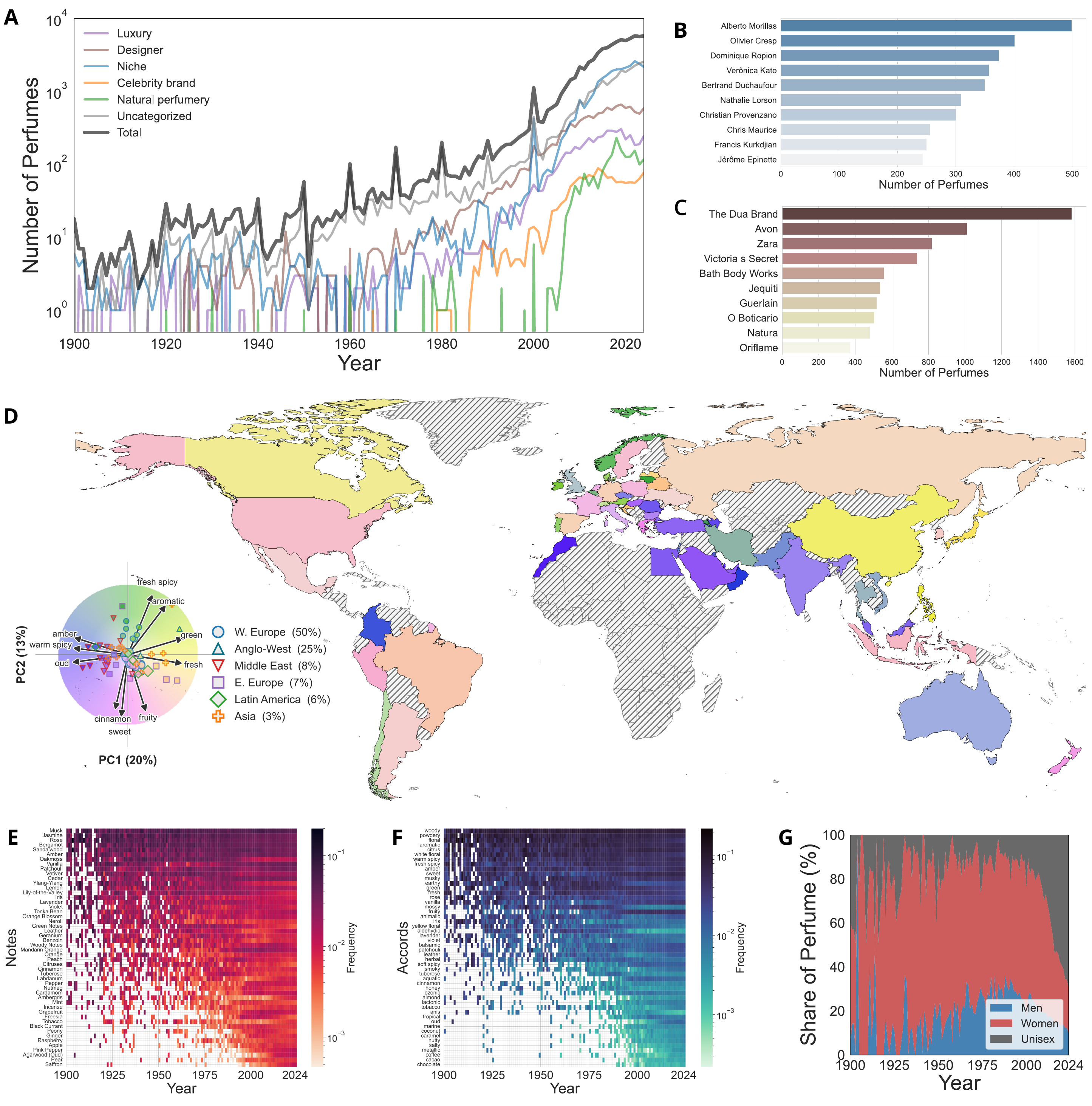} 
	\caption{\textbf{Overview of the Perfume dataset.} \textbf{(A)} Annual number of perfume releases from 1900 to 2024 (logarithmic scale on the y-axis). Different colors indicate perfume categories: luxury houses (purple), designer perfumes (brown), niche perfumes (blue), celebrity-branded perfumes (orange), and natural perfumery (green), together with an unclassified group (gray) and total releases across all categories (black). \textbf{(B)} Number of commercial perfumes created by the top ten most productive perfumers. \textbf{(C)} Total perfumes released by the top ten perfume brands. \textbf{(D)} Global map colored by each country's accord profile of the top 40 accords (PCA of accord usage relative to the global average). Similar colors mean similar profiles, and gray hatching marks countries with fewer than 30 perfumes. The inset is the biplot of the same space, with points colored as on the map, shaped and outlined by region and sized by percentage of global releases, with arrows for the top 10 leading accords. The legend lists each region's release percentage share. \textbf{(E-F)} Frequency of the top 50 perfume notes \textbf{(E)} and perfume accords \textbf{(F)}, represented on logarithmic scales. \textbf{(G)} Share of perfume releases (\%) categorized by intended gender: men (blue), women (red), and unisex (gray).}
	\label{fig1} 
\end{figure}

We assembled a comprehensive perfume dataset from three databases, namely Fragrantica, Parfumo, and Aromo, and we additionally obtained a perfume ingredient list from the Fragrance Ingredient Glossary (see \hyperref[Methods]{Methods}). Together these platforms contribute a dataset comprising 169,759 perfumes, 8,332 perfume brands, 3,050 individual perfumers, and approximately 25.9 million crowd-sourced votes on perfume notes (i.e., distinct scents perceived in a perfume). The summary statistics of all datasets are presented in Table~\ref{tab:dataset_comparison}. The cross-database comparison analysis shows that these datasets agree closely on composition, release history, and note usage (see \hyperref[Supplementary information]{Supplementary information} and Figure \ref{fig:S1}). Thanks to Fragrantica's substantially larger user base and international traffic, it yields far richer and more complete data coverage than the other two sources, in particular the crowd-sourced data and the resemblance annotations on which several of our analyses depend. Thus, in this study, we present the analysis from Fragrantica in the main text, and we use Parfumo and Aromo for validation, presenting those results in the \hyperref[Supplementary information]{Supplementary information}. 

The crowd-sourced votes from Fragrantica assess multiple qualitative dimensions, including longevity (i.e., how long the aroma lasts), sillage (i.e., the trail of scent a perfume leaves in the air), gender association (i.e., whether a perfume is perceived as masculine, feminine, or unisex), price value, overall rating (i.e., how much users liked the scent), and resemblance annotations (i.e., what other perfumes a given one brings to mind). The dataset combines both manufacturer-reported information and user-generated content. For example, compositional attributes such as perfume notes and gender classification are typically reported by the brands themselves and subsequently receive crowd-sourced votes from users, whereas attributes such as longevity, sillage, and price value are derived solely from crowd-sourced input. 

Our dataset contains a single perfume from 1533, no records through the seventeenth century, and only sporadic entries across the eighteenth and nineteenth centuries before releases accelerate in the twentieth century. Thus, we focus our analysis on the period from 1900 to 2024. The annual number of perfume releases from 1900 to 2024 is shown in Figure \ref{fig1}A. Different colors indicate distinct segments of the perfume industry, including luxury houses (purple), designer perfumes (brown), niche perfumes (blue), celebrity-branded perfumes (orange), and natural perfumery (green), together with an unclassified group in our dataset (gray) (Figure \ref{fig1}A). Here, ``Designer'' denotes perfumes issued by fashion and lifestyle houses (e.g., Hugo Boss, Calvin Klein), whereas ``Luxury'' refers to a set of high-end and prestigious brands (e.g., Chanel, Dior, Creed). Total releases across all segments are shown as a thick solid line (black). Luxury, designer, and niche perfumery date back to the early twentieth century, whereas celebrity-branded perfumes began to increase markedly in the late 1980s, and natural perfumery entered a steady rise around 2000 after appearing only sporadically in the early twentieth century. These changes reflect broader sociocultural shifts and market transformations. In particular, natural perfumery, which emphasizes naturally derived ingredients, has seen a drastic growth that aligns with more recent consumer trends toward sustainability \cite{dwivedi2025evolution}. In contrast, luxury, designer, and niche perfumes show sustained long-term growth, which reflects their consistent presence in the perfume market. 

Perfume releases in the last century grew in two waves. The first inflection point near 1918 established a carrying capacity of about 20 releases per year (a plateau spanning around 1926–1979), and a second inflection point around 2014-2018 drove expansion toward a carrying capacity of roughly 6,000 new perfumes per year (Figure \ref{fig:S2}). These transitions separate three eras which we refer to as, a Classic Perfumery era of gradual growth, a Stagnation period, and an Industrial Perfumery era of rapid, large-scale expansion. The latter two align with records of fragrance environmental deposition \cite{vecchiato2020great} and consolidation in the fragrance industry \cite{jones2010beauty}.

To understand artisanship and industry constraints, we analyzed the most prominent perfumers and brands. The ten most prolific perfumers are shown in Figure \ref{fig1}B. Alberto Morillas is the most productive master perfumer, having created nearly 500 commercial perfumes. The output of the top ten brands is shown in Figure \ref{fig1}C. The Dua Brand, known for high-volume, affordable clones of designer perfumes, ranks highest with 1,584 perfumes, followed by mass-market brands Avon, Zara, and Victoria's Secret. The only high-end designer brand that appears in this list is Guerlain.

To characterize the interplay between geography and scent, we summarized each country by profile of olfactory accords (i.e., an emergent feature reflecting a perfume's overall scent impression) in the perfumes it produces, and projected these profiles onto a global map (Figure \ref{fig1}D). We expressed each country's accord frequencies as deviations from the global average using log location-quotient, reduced them to two dimensions by principal component (PC) analysis, and colored each country by its position in this accord space. PC1 and PC2 capture 20\% and 13\% of the variance, respectively. Most countries, including those of Western Europe, the Anglosphere, Latin America, and much of Asia, cluster near a global mainstream built on fresh, fruity, and sweet accords. The clearest departure from this pattern is the Middle Eastern countries, whose perfumes over-represent oud, amber, and warm-spicy accords, consistent with longstanding oriental perfume traditions \cite{stamelman2006perfume}. In terms of output, Western Europe accounts for roughly half of the catalogued perfumes and the Anglosphere for about a quarter. In sum, geographic region explains a modest but significant share of national scent profiles of 13\% for notes and 17\% for accords (see \hyperref[Supplementary information]{Supplementary information} and Figure \ref{fig:S3} for regional scent profiles).

Next, we analyzed the temporal evolution of perfume notes and accords (Figure \ref{fig1}E-F). The heatmap in Figure \ref{fig1}E shows the annual frequency of the top 50 notes. Among these, Musk, Jasmine, Rose, Bergamot, Sandalwood, and Amber stand out for their consistent prominence across decades. The sparse data before 1950 reflect the smaller number of perfumes from that period. In later decades, especially after the 1990s, top notes shifted dynamically (e.g., oakmoss and ylang-ylang declined while vanilla and cedar rose). Regulatory restrictions imposed by the International Fragrance Association (IFRA) partly drove these dynamics. For example, oakmoss, a known skin sensitizer, was progressively restricted, with its maximum level lowered to 0.1\% of the finished product by 1997 \cite{marie2011regulatory}. 

A similar heatmap for perfume accords is shown in Figure \ref{fig1}F. The top ten accords over time include Woody, Powdery, Floral, Aromatic, Citrus, White Floral, Warm Spicy, Fresh Spicy, Amber, and Sweet. Particularly, Fruity accords in the 1990s experienced a continued growth into the 2000s, while Earthy, Mossy, and Aldehydic accords show a pronounced decline. The Aldehydic accord became one of the most widely used after its employment in Chanel No 5, which influenced modern perfumery in the early 20th century \cite{turin2008perfumes} and its decline in the early 1990s represents a cultural pivot in perfumery aesthetics. 

Finally, we analyzed the temporal trends in perfume gendering which is aligned with broader cultural movements toward fluid identity expression \cite{bauman2013liquid}.  The annual market shares of perfumes categorized as female, male, or unisex are shown in Figure \ref{fig1}G. Female-marketed perfumes dominated for most of the 20th century. However, since the late 1990s, a notable rise in unisex perfumes suggests a gradual shift toward gender-neutral scent marketing. Interestingly, the gender association of many notes and accords has shifted, but not toward neutrality (Figure \ref{fig:S4}). Among gendered perfumes alone, the core of gender-coding still holds, where tuberose, iris, rose, and jasmine stay feminine, while sage and leather stay masculine. The net movement is mainly from feminine to masculine (around 61\% of the significant changes). Some descriptors that cross the neutral line include styrax, incense, spicy notes, vetiver, oakmoss, lemon, and petitgrain for notes, and cinnamon, warm spicy, earthy, and patchouli for accords (see Figure \ref{fig:S4}).

\subsection{Perfume Complexity and Novelty}

\begin{figure}[!htbp]
	\centering
	\includegraphics[width=1\textwidth]{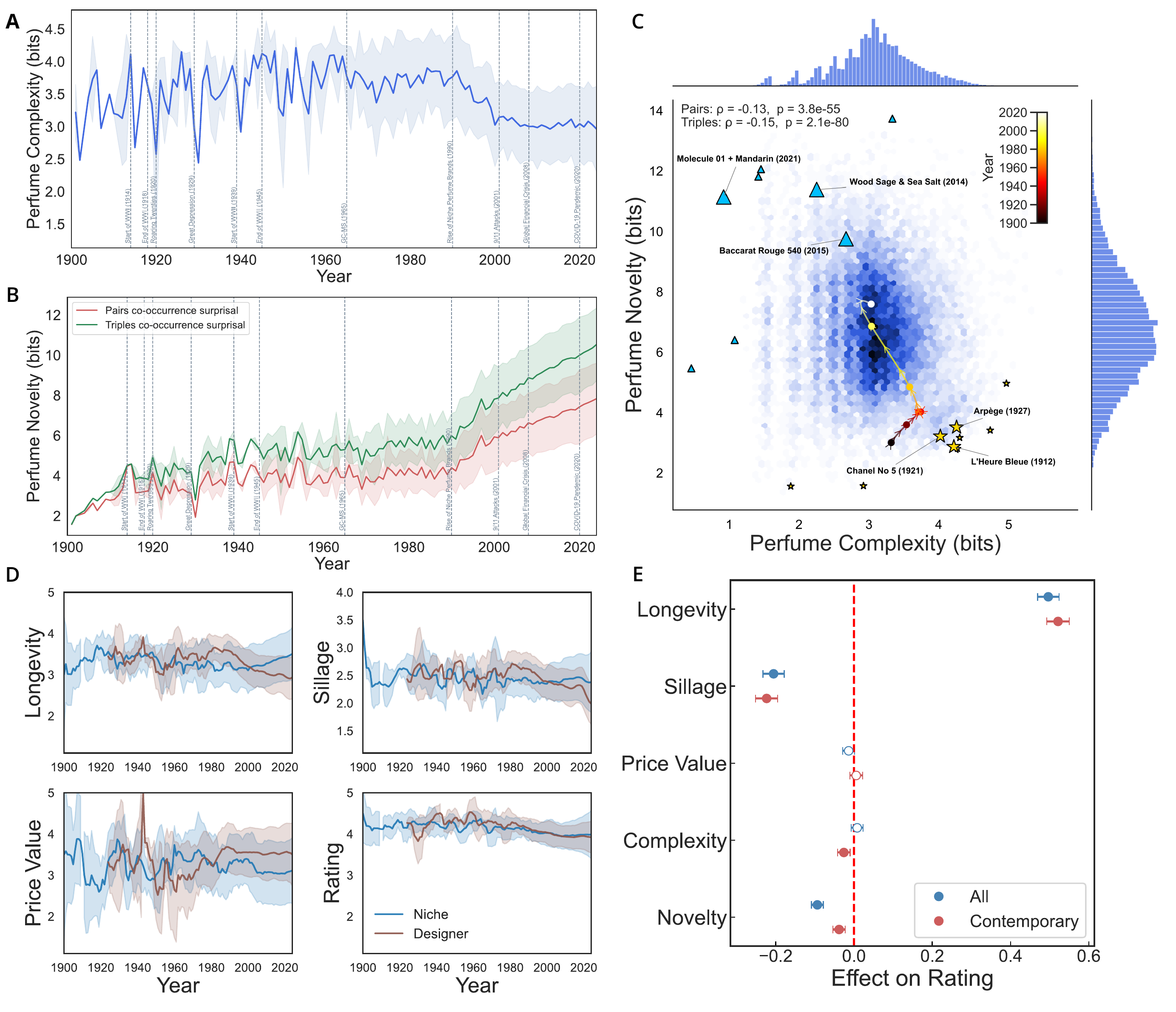} 
	\caption{\textbf{Trends in perfume characteristics.} \textbf{(A–B)} Annual mean perfume complexity \textbf{(A)} and perfume novelty \textbf{(B)} from 1900 to 2024. Solid lines are yearly means and shaded bands are standard deviation. Complexity is the Shannon entropy of a perfume's note distribution in bits, and novelty is the note co-occurrence surprisal against all earlier perfumes in bits, shown for both note pairs and triples. Vertical dashed lines indicate major social, cultural, and technological events, from left to right: the start and end of WWI (1914, 1918), the Roaring Twenties (1920), the Great Depression (1929), the start and end of WWII (1939, 1945), the commercial introduction of gas chromatography/mass spectrometry (GC-MS, 1965), the rise of niche perfume brands (1990), the 9/11 attacks (2001), the global financial crisis (2008), and the COVID-19 pandemic (2020). \textbf{(C)} Joint distribution of complexity and pair novelty. The coloured line traces the decade-averaged trajectory over time, with colour indicating year, and the inset reports the Spearman correlation between complexity and novelty. Highlighted perfumes identify the two Pareto fronts. Gold stars denote the high-complexity, low-novelty frontier, namely Chanel No~5 (1921), Guerlain's L'Heure Bleue (1912), and Lanvin's Arpège (1927). Blue triangles denote the low-complexity, high-novelty frontier, namely Wood Sage \& Sea Salt (2014), Baccarat Rouge 540 (2015), and Molecule 01 + Mandarin (2021). \textbf{(D)} Annual mean longevity, sillage, price value, and user rating for niche brands (blue) and designer brands (brown), lines are 5-year rolling means and shaded bands are standard deviation. \textbf{(E)} Coefficient plot of standardized effect sizes, with 95\% confidence intervals, from OLS regressions of average user rating on perfume characteristics (complexity, novelty, longevity, price value, and sillage ) for all perfumes (blue) and contemporary perfumes released since 2000 (red). Novelty is the mean of the pair and triple co-occurrence surprisal. P-values are corrected using the Benjamini–Hochberg false discovery rate. Filled markers denote significant associations and open markers non-significant ones.}
	\label{fig2} 
\end{figure}

We model the complexity of a given perfume as the Shannon entropy of its notes probability (see \hyperref[Methods]{Methods}). Figure \ref{fig2}A shows the temporal trends in perfume complexity, where the solid line corresponds to the mean perfume complexity per year, computed as the average Shannon entropy of all perfumes of a given year, and the shaded area shows its standard deviation. Distinct phases in the dynamics emerge from the data. Initially, between 1900 and 1960, the complexity trajectory exhibits significant fluctuations, primarily driven by the relatively low number of perfumes. These fluctuations begin to diminish around mid 1960s, coinciding with an exponential growth in new perfume releases (Figure \ref{fig1}A). After a phase of relative stability in 1965-1990 with an average entropy of around 3.7 bits, a pronounced decline in perfume complexity occurs abruptly around 1990, that follows a transition to another stable period spanning over two decades from 2000 to 2024 with an average entropy of around 3.0 bits. We observe the same declining trend in a uniform-weight variant of complexity in the Parfumo dataset. The complexity estimate is unreliable for Aromo, as only 14.5\% of its catalog records both a release year and notes, versus around 82\% for Fragrantica and 38\% for Parfumo with both records (Figure \ref{fig:S5}A).

Next, we measure novelty at the level of note combinations. We quantify a perfume's novelty as the surprisal of its co-occurring note pairs and triples, measured against the frequency of each combination among perfumes from earlier years. A combination that was common among earlier perfumes is unsurprising and adds little novelty, whereas rare combinations are more surprising and thus more novel. Each combination is weighted by the prominence of its constituent notes, and averaging these weighted surprisals yields a single novelty value, in bits (see \hyperref[Methods]{Methods}). The resulting temporal dynamics are shown in Figure \ref{fig2}B. Novelty rose steadily and strongly across the entire period, for both note pairs and triples, climbing for pairs from about 2.3 bits in the early 1900s to 7.6 bits in the 2020s, with the steepest gains from the 1990s onward. The same trends hold in Parfumo and more weakly in Aromo, under an unweighted variant of novelty measures (see Figure \ref{fig:S5}B).

We also recover novelty change with an independent semantics-based measure, by representing each note by its Word2Vec embedding \cite{mikolov2013efficient}. Semantic novelty likewise rises from the 1990s onward for both Fragrantica and Parfumo (see \hyperref[Supplementary information]{Supplementary information} and Figure \ref{fig:S5}C). Because the semantics-based measure does not rely on combinatorial counts and is invariant to palette size, its agreement with the co-occurrence result indicates that the rise in novelty is a genuine pattern rather than a side effect of palette expansion.

Complexity and novelty tend to move in opposite directions at the population level, particularly in the data-rich period after 1965. This indicates that while perfumes became structurally simpler, they draw on increasingly rare note combinations. The palette of perfumers has been steadily growing over the last century. The cumulative number of distinct notes across all three databases, and of distinct odor descriptors in the Fragrance Ingredient Glossary (FIG), has expanded by more than an order of magnitude since 1900 (Figure \ref{fig:S6}). In addition, complexity declines similarly across the top notes (i.e., high-volatility molecules, emitted first), middle notes (i.e., mid-volatility, emerging after a while), and base notes (i.e., low-volatility, lasting for hours) of the perfume pyramid, whereas for novelty the top and middle notes contributed more than the base notes to the total novelty (see Figure \ref{fig:S7}).

The joint distribution of perfume complexity and novelty is shown in Figure \ref{fig2}C. Across perfumes, complexity and novelty are weakly related, where more complex perfumes tend to be less novel, with negative association that is small in magnitude but statistically significant (Spearman correlation $\rho = -0.13$ for pairs and $-0.15$ for triples). The year-colored trajectory traces the mean position of perfumes through the novelty-complexity plane over time. From about 1900 to 1960, perfumes moved toward both higher complexity and higher novelty. After mid-century, the path reverses in complexity while continuing to climb in novelty and perfumes became progressively simpler yet increasingly unconventional, which leads to a tradeoff between complexity and novelty. The Pareto fronts in Figure \ref{fig2}C define the boundaries of this space. Its conventional, low-novelty frontier shown by gold stars is occupied by complex early-twentieth-century classics built on well-established notes, such as Chanel No~5 (1921), Guerlain's L'Heure Bleue (1912), and Lanvin's Arpège (1927). The high-novelty frontier shown by blue triangles is occupied by recent minimalist releases assembled from rare note combinations, such as Wood Sage \& Sea Salt (2014), Baccarat Rouge 540 (2015), and Molecule 01 + Mandarin (2021).

Next, we examined how various perfume attributes, including longevity, sillage, price value, and user ratings have changed over time (Figure \ref{fig2}D). Longevity has fluctuated over the past century, with a modest increase after 2000 for niche brands but a decline for designer perfumes (upper left panel, Figure \ref{fig2}D). Sillage shows a similar pattern, with niche brands surpassing designer brands, although the overall trend is a gradual decline in sillage, for designer perfumes (upper right panel, Figure \ref{fig2}D). The price value is rated higher for niche perfumes, so designer perfumes are judged to be more expensive (lower left panel, Figure \ref{fig2}D), while average user ratings remain comparable between niche and designer brands (lower right panel, Figure \ref{fig2}D).

Examining how perfume attributes relate to user ratings for all perfumes and for contemporary perfumes (2000-2024) (see \hyperref[Methods]{Methods} and Figure \ref{fig2}E) reveals several patterns. Longevity shows by far the strongest positive association, indicating that consumers prefer perfumes that last longer. Sillage, by contrast, is negatively associated with ratings, a counterintuitive result consistent with the gradual decline in sillage over time, in particular for designer perfumes. Perceived value for money has no significant effect. Complexity, though not significant across all perfumes, becomes significantly negative among contemporary perfumes, and  novelty (the mean of pair and triple surprise) is negatively associated with ratings in both groups. These results indicate that users favor simpler and less experimental perfumes.

\subsection{Imitation and Cultural Selection in Perfumes}
Perfume offers a rare window onto cultural copying. Perfume flankers (i.e., a spin-off of successful perfume from the same brand), or clones (i.e., replicate of high-end perfumes from other brands) are announced and debated in public. 
%\textcolor{red}{Since the commercial introduction of gas chromatography/mass spectrometry (GC-MS) in 1965, it became possible to accurately decode perfume formula and produce clones. The effect of GC-MS became significant ?? years later as...} 
Fragrantica organizes this information on the platform, and perfume wearers crowd-source it by voting. This leaves a direct trail of validated imitation that is invisible in most other creative domains. We exploited this by using a crowd-sourced resemblance annotation record to construct time-directed perfume lineage network. 

\begin{figure}[!htbp] 
	\centering
	\includegraphics[width=1\textwidth]{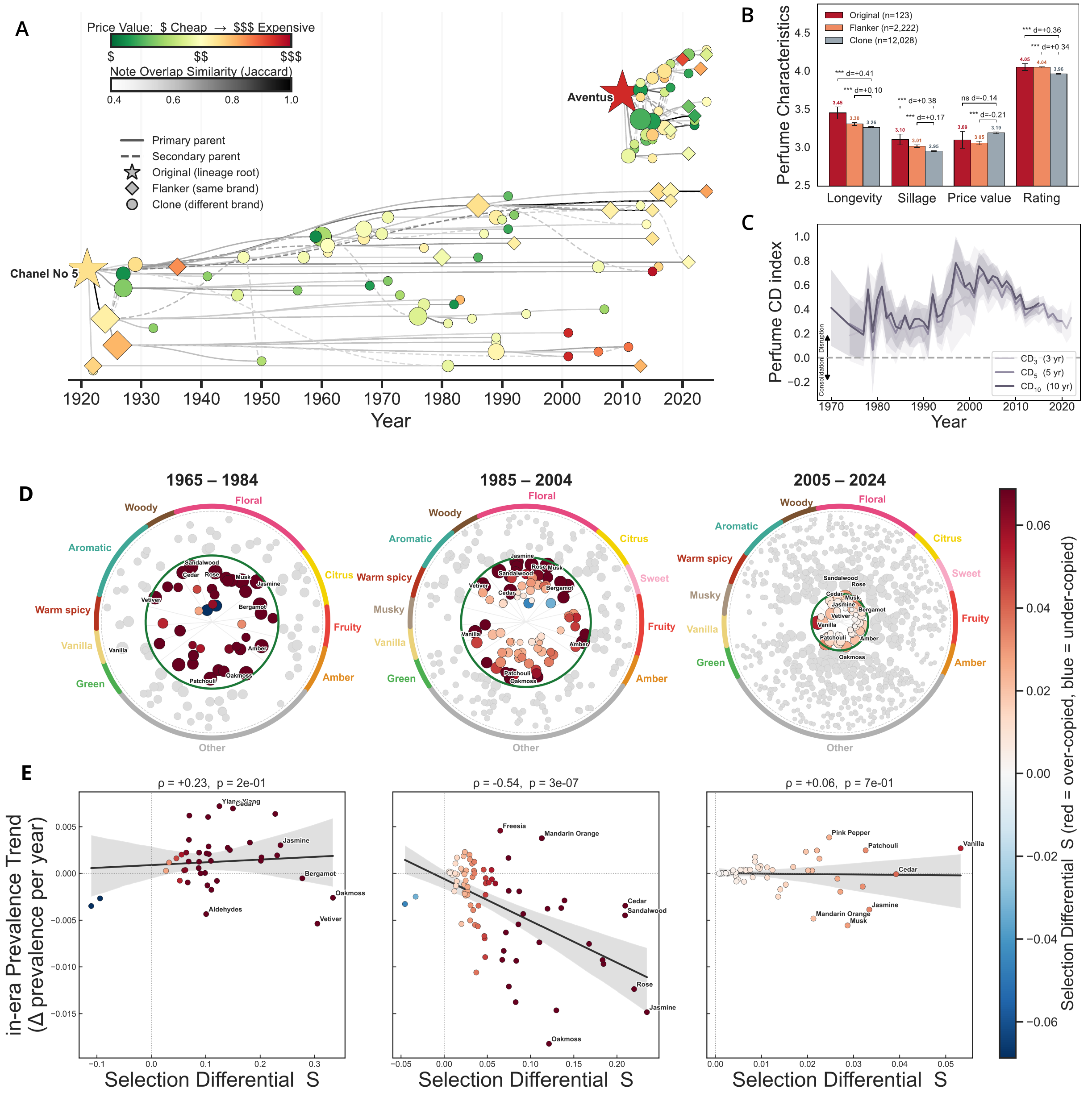} 
	\caption{\textbf{Imitation as a selective force in the cultural evolution of perfume.} \textbf{(A)} The two largest copy lineage networks, Chanel No 5 and Aventus, shown on a release-year axis. Solid curves link each perfume to its primary parent, dashed to a secondary parent. Marker shape denotes Original (star), Flanker (diamond) or Clone (circle), and the size corresponds to the number of imitators, fill colour shows the price value (green to red denotes transition from great value to expensive) and edge shade the note overlap (Jaccard index). \textbf{(B)} Mean value of crowd-rated longevity, sillage, price value and rating for Originals ($n = 123$), Flankers ($n = 2{,}222$) and Clones ($n = 12{,}028$). Error bars are 95\% CI, and p-value is calculated using two-sided Mann-Whitney $U$, Cohen's d is reported as the effect size. \textbf{(C)} CD index on the copy lineages network over 3, 5 and 10-year windows (shaded band shows 95\% CI). \textbf{(D)} Per-era selection map, where notes arranged by accord sector. Coloured notes show significant selection after FDR correction, but gray points are not significant and drifting notes. The green ring's area as a fraction of the total disk is the fraction of notes under selection. \textbf{(E)} For notes with significant selection, the in-era prevalence trend against selection differential, over-copying and rising use are negatively coupled in 1985-2004, but is not significant relation with in other eras. Data points in \textbf{(D-E)} are painted by their selection differential $S$ (red indicates over-copied and blue indicates under-copied).}
	\label{fig3} 
\end{figure}

The resulting lineage network can be treated as an evolutionary network of perfumes where parent perfumes (Originals) sets the ground for the creation of flanker or clones. Resemblance annotations capture genuine crowd-sourced perceptual similarity and compositional clone rather than a loose association. In our dataset and across highly-voted perfumes (see \hyperref[Methods]{Methods}) we find $169{,}991$ perfume pairs with resemblance annotation record. These resembling pairs share far more notes than what would be expected from a null model, with a mean Jaccard index of $J=0.17$ versus $J=0.09$ for random pairs (see Figure \ref{fig:S8}). Here, we take the top five percent of this distribution ($J^{\ast} \geq 0.39$) as a data-driven threshold, and by iterating the copy relation backward in time for each perfume, we construct a lineage network. Chanel No 5 and Aventus by Creed are the two biggest genealogies that emerge in our computational pipeline ( Figure \ref{fig3}A). Chanel No 5 seeds lineages to $68$ perfumes spanning a century, and Aventus is a pillar for creation of $36$ perfumes in just a single decade. Other most-copied scents include iconic perfumes such as Ex'cla-ma'tion by Coty, Aloha Tiare by Comptoir Sud Pacifique, Gentleman (1974) by Givenchy, and Fracas by Robert Piguet (see Figure \ref{fig:S9}). Two patterns are worth noting. First, most copying happens across brands; among the 2,563 descendants in all trees, 62.6\% are cross-brand clones and only 37.4\% same-brand flankers. Second, lineages are not isolated; 28.6\% of perfumes have a second parent alongside their primary one, which resembles hybridization in biological evolution \cite{taylor2019insights}. From these data, we estimate that clones make up at least 1.0–11.7\% of perfumes in the market, depending on the note-overlap threshold (see Figure \ref{fig:S10}).

We then asked whether Originals differ in quality from the perfumes that copy them. By comparing Originals with Flankers and Clones on four crowd-rated attributes, we find that Originals are judged longer-lasting (longevity $3.45$ versus $3.26$), left a stronger scent-trail (sillage $3.10$ versus $2.95$), and are better overall (rating $4.05$ versus $3.96$) than Clones, with Flankers sitting consistently between the two (Figure \ref{fig3}B). The one exception is price value, on which Clones are rated slightly higher which is consistent with copies competing as cheaper alternatives, although this contrast is not statistically significant.

Next, we examined whether perfumes have grown more or less disruptive in breaking from accepted stylistic norms, by applying the consolidation-disruption (CD) index~\cite{park2023papers} to the lineage network (see \hyperref[Methods]{Methods}). On average the field is disruptive  with mean CD of $0.18$, $0.21$ and $0.26$ over 3, 5, and 10 year windows, but its trajectory is not flat (Figure \ref{fig3}C). We observe that disruption rises through the late twentieth century, peaks around $2000$, and then declines steadily toward the consolidation regime over the last two decades. The rise in disruption through the late twentieth century coincides with the rise in novelty (Figure \ref{fig2}B), suggesting the two may be linked.

Finally, to measure selection at the level of individual notes, we computed the selection differential of the notes using the Price equation~\cite{price1970selection} across three successive 20-year eras, namely a pre-complexity-transition period (1965–1984), the complexity-transition period (1985–2004), and a post-complexity-transition period (2005–2024) (see \hyperref[Methods]{Methods} and Figure \ref{fig3}D). The fraction of notes under significant copy-selection drops sharply across eras, from 36\% (1965-1984) and  34\% (1985-2004) to just 7\% (2005-2024), and the selection differential also weakens among the selected notes (S=+0.11, +0.06, +0.01). The area enclosed by the green ring as a fraction of the total disk in Figure \ref{fig3}D is equal to the fraction of selected notes. The shrinkage of selected notes suggests a shift from selection toward drift in recent decades. Furthermore, we validated selection against the prevalence trend of the notes as an independent quantity (Figure \ref{fig3}E). The association between being over-copied and a rising trend is weak and non-significant in 1965-1984 and 2005-2024. In the complexity transition period in 1985-2004, however, this coupling is strong, negative, and statistically significant ($\rho=-0.54$, $P=3\times10^{-7}$), with the era's most over-copied notes effectively falling out of use. During this period, IFRA imposed varying levels of restriction on notes such as oakmoss \cite{marie2011regulatory}.

\subsection{Social Contagion of Style in Perfumer's Collaboration Networks}

\begin{figure}[!htbp] 
	\centering
	\includegraphics[width=1\textwidth]{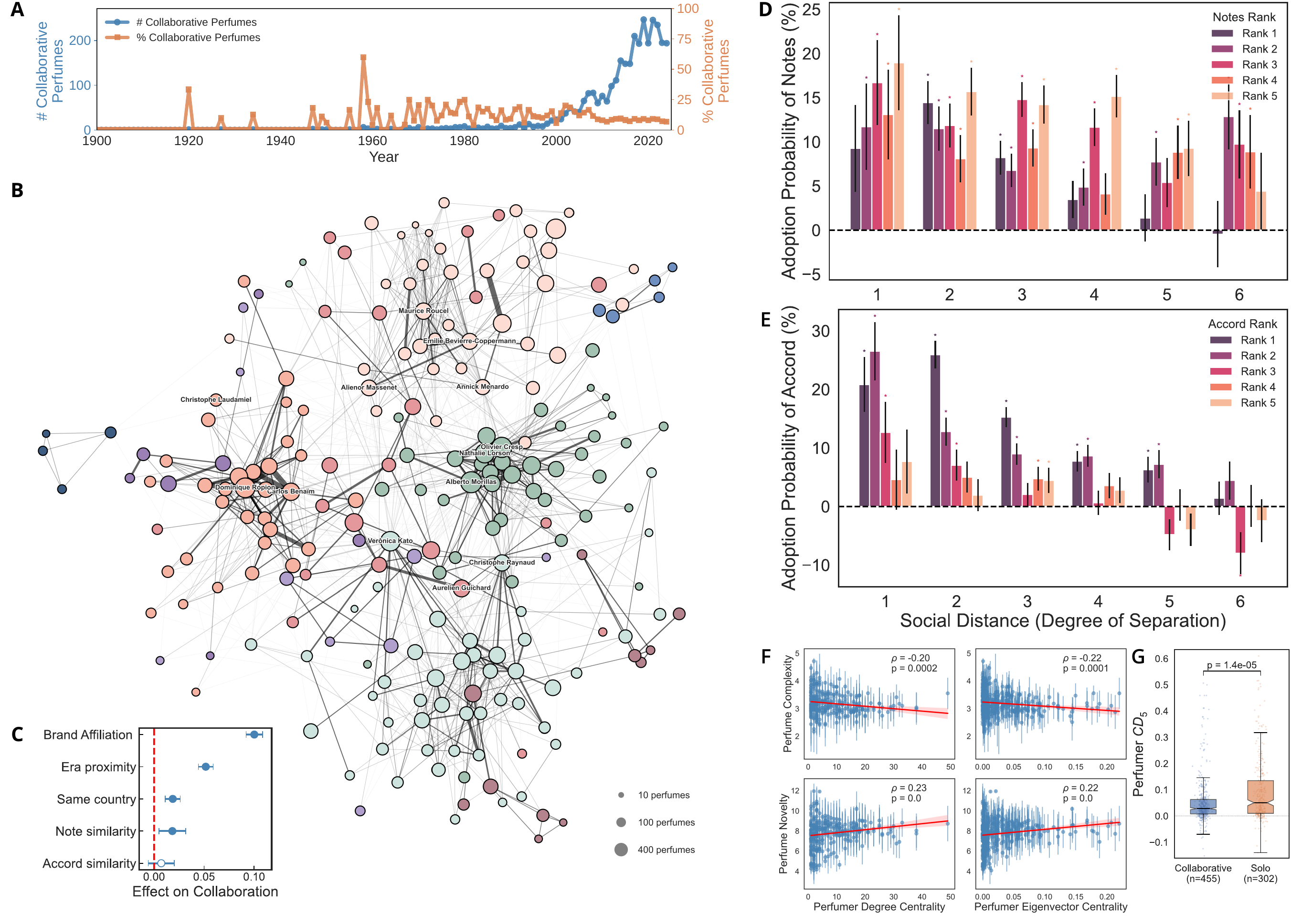} 
	\caption{\textbf{Collaboration and the social transmission of perfumers' styles.}
    \textbf{(A)} Annual count (left axis, blue) and fraction of collaborative perfumes (right axis, orange). \textbf{(B)} The 3-core of the largest cluster of the collaboration network. Node size is the number of perfumes made and node colour the collaboration community. Edges are black within a community and gray between, with thickness proportional to co-created perfumes. Labels mark the leading perfumers. \textbf{(C)} Standardized regression coefficients (95\% CI) for five pairwise predictors of collaboration (shared brand, era proximity, note similarity, same country, and accord similarity), each measured from the two perfumers' work before their first joint perfume, the red dashed line indicates no effect and open circles mean not significant. \textbf{(D-E)} Relative change in the probability that a perfumer adopts five top-ranked descriptors of a note \textbf{(D)} or an accord \textbf{(E)} from a network-connected source. \textbf{(F)} Mean perfume complexity (top) and novelty (bottom) as a function of degree (left) and eigenvector centrality (right) of a perfumer, respectively. Markers are per-perfumer means and error bars are standard deviation, red lines are regression fits with 95\% CI. \textbf{(G)} Mean $\mathrm{CD_5}$ index for perfumers who ever collaborated versus always-solo
    (p-value is calculated with two-sided Mann-Whitney $U$).}
	\label{fig4} 
\end{figure}

Collaboration is ubiquitous in many human activities and in the making of cultural artifacts, yet the authorship norms also vary. Some are predominantly collective, such as music, theater, film. Others carry a strong tradition of individual authorship, such as painting, sculpture, poetry, and literature. Perfume creation was predominantly a solitary craft throughout perfumery history, compositions attributed to a single perfumer (also known as the \emph{nose}), with very few recorded instances of collaboration between perfumers (such as Knize Ten in ~1920). In this respect, perfume has historically resembled painting more than music. Across the era covered by our data, collaboration has remained a minority practice, with collaborative perfumes making up roughly 8-15\% of credited perfumes in any given year. The absolute number of collaborative perfumes, however, has grown sharply since the late 1990s (Figure \ref{fig4}A), but this tracks the overall expansion of perfume output rather than a rising tendency to collaborate. The collaborative share has declined, from around 13\% through the 2000s to around 8\% in the 2010-2020.

Using records of collaborative perfumes, we constructed the perfumer collaboration network. In Figure \ref{fig4}B, we display the 3-core of the largest cluster of the collaboration network, comprising 219 of its 500 perfumers. Each node represents a perfumer, and each edge indicates that the connected perfumers have co-created at least one perfume. The thickness of the edges corresponds to the number of perfumes created together, the size of each node is proportional to the number of perfumes the perfumer has created, and the color of each node denotes the collaboration community to which the perfumer belongs. Collaboration clusters are identified using the Louvain algorithm \cite{blondel2008fast}. The most-connected perfumers are labeled by name. Smaller clusters of the collaboration network are shown in Figure \ref{fig:S11}. 

To identify what brings perfumers together, we tested whether pre-existing affiliation or stylistic similarity between two perfumers predicts that they subsequently collaborate, using only the perfumes each had created before their first joint release (see \hyperref[Methods]{Methods} and Figure \ref{fig4}C). Shared brand affiliation was the strongest driver, followed by proximity in activity era, sharing a country, and pre‑existing note similarity, whereas pre‑existing accord similarity was not significant. This implies that collaborations in perfumery primarily form through institutional affiliation rather than prior strong stylistic similarity. 

We then tested for social contagion: do perfumers adopt the stylistic features of their collaborators? For each ordered pair of perfumers at a given social distance, we computed the relative increase in the probability that a perfumer adopts a note or an accord  that a network-connected source had used before the two became connected (see \hyperref[Methods]{Methods}). To separate contagion from pre-existing similarity (i.e., homophily in the interactions), we compared the probability of note usage against a matched null in which the source is replaced by a random non-collaborator matched both on prior stylistic overlap with the focal perfumer and on whether they shared a brand with them. We performed 5{,}000 simulations and the corresponding null is shown by the black dashed lines at zero in \ref{fig4}D-E. The relative increase in note adoption is highest at the first degree of social distance (i.e., direct collaborators), with an average value of $13.94$ for the top 5 notes (Figure \ref{fig4}D). This declines to the average values of $12.32$, $10.65$, $7.85$, $6.52$, and $7.08$ from the second to the sixth degree of separation, respectively. A similar but slightly weaker pattern is observed for accord adoption (Figure \ref{fig4}E), with the average values of $14.46$, $10.53$, $7.09$, $4.66$, $0.98$, and $-0.89$ from the first to the sixth degree of separation. The contagion is asymmetric and driven by prestige difference, where it flows more strongly from high-prestige to low-prestige perfumers. (Figure \ref{fig:S12}). 

We next examined how complexity and novelty relate to positions within the collaboration network. For each perfumer, we calculated the average complexity and novelty scores of their creations, and assessed their associations with collaboration network properties, including perfumer degree and eigenvector centrality. As shown in  Figure \ref{fig4}F, we observe statistically significant negative Spearman's rank correlation between mean perfume complexity for both perfumer degree and eigenvector centrality. Additionally, we found statistically significant positive Spearman's rank correlations  between mean perfumer novelty and perfumer degree and eigenvector centrality. 

Finally, we asked whether collaboration relates to disruptiveness of a perfumer's work, quantified by $\mathrm{CD_5}$. Perfumers who had ever collaborated were significantly less disruptive than those who always worked solo with median $\mathrm{CD_5=+0.028}$ versus $\mathrm{CD_5=+0.050}$, respectively (Figure \ref{fig4}G). This suggests that collaboration is associated with more consolidating rather than disruptive creations, similar to the effect of team size on the CD of science and technology \cite{wu2019large}. This difference is robust across the $\mathrm{CD_3}$ to $\mathrm{CD_{10}}$ windows (see Figure \ref{fig:S13}).

\section{Discussion}\label{Discussion}

Using the largest dataset of perfumes, we find that perfumes evolve in their style and structure in discernible ways. Geography partially encodes their global production pattern, suggesting that culture partially encodes odor pleasantness and olfactory perception in general \cite{majid_human_2021,arshamian_perception_2022,drnovsek2025demographic}. From 1900 to 1965 we observe large fluctuations in both perfume complexity and novelty. Some of the rises and falls in complexity coincide with large-scale global events. For instance, the onset of both World War I and II coincides with the decline in complexity and novelty. This could be the result of the disruption in supply chains of perfumery ingredients. On the other hand, in the 1920s we observe a steady increase in complexity, possibly due to social and economic effects of the Roaring Twenties on the fashion market. Despite these evidently meaningful associations between global events and fluctuations in the complexity and novelty, due to the small sample size before 1965 (less than 2\% of perfumes), these results could also be influenced by numerical uncertainties. However, after 1965, we see more stable trends in both complexity and novelty, which follow a transition from complex to minimal perfumes, and a monotonic increase in perfume novelty. The transition starts in the early 1990s and it enters the minimal perfume state in the early 2000s and continues to rise in the exploration of novel combinations.

Disentangling all the mechanistic forces that drive large-scale transitions in perfumery is difficult. Much happened in the second half of the twentieth century that may have shaped this dynamic, from consolidation and mergers in the beauty and fragrance industry \cite{jones2010beauty} to the rise of regulation in the fragrance industry \cite{marie2011regulatory}, such as IFRA's restrictions, to the rise of niche perfume brands and online marketplaces. 

However, we can still discuss some potential scenarios. First, perfume houses and brands can be expected to pursue market revenue. Optimizing for market revenue might have directly caused a transition from longer perfume formulas to shorter ones with fewer perfume ingredients. Second, in the twentieth century, minimalism emerged from avant-garde circles and became mainstream in several cultural domains such as architecture and design (1920), fine art (1960), linguistics (1970), and tech design (2009) \cite{vaneenoo2011minimalism}. The transition to minimalism in perfumery may be a natural extension of the minimalism movement into perfume artisanship. Third, by the late 1990s and early 2000s, online marketplaces began to gain traction, which in turn helped independent perfumers gain prominence, both by having easier access to perfume ingredients and by selling their products online. Working with short formulas is easier practically for independent perfumers, and this may have been one of the forces driving the transition. Last, we find that more connected perfumers with higher centrality in the collaboration network (Figure \ref{fig4}F) tend to create less complex yet more novel perfumes. This may imply a spillover effect, whereby more central perfumers influence peripheral and potentially junior perfumers to move toward minimalism in perfumery (Figure \ref{fig:S12}). One of the outstanding examples is the influence of Edmond Roudnitska, known for his short formulas, on the new generation of perfumers \cite{rudnitska2000parfum}. These observations provide potential mechanistic causes of the observed stylistic transitions in Figure \ref{fig2}A-B. 

The significant negative associations between perfume complexity and novelty (Figure \ref{fig2}C) points to a plausible exploration mechanism in which constraints on perfume ingredient use drive innovation through a natural feedback loop. Exogenous regulatory restrictions on ingredient use offer a natural experiment to test this. For example, IFRA's restrictions on oakmoss drove affected perfumes to become both simpler and markedly more novel, with the effect strengthening as the restriction became more severe (Figure \ref{fig:S14}). This provides evidence that external constraint drives exploration, consistent with the idea that the constraints behind the minimalist transition pushed perfumers toward exploring novel combinations to help their creations stand out.

The perfume genealogical network lets us study perfumery as a population of cultural variants under selection \cite{newberry_measuring_2022,price1970selection} rather than a sequence of isolated creations. Copying perfumes is genealogical and reticulate, where perfumes descend from a few lineages and fuse across brands into a network that behaves like a recombination space \cite{youn_invention_2015}. As expected, Clones compete as cheaper alternatives to their high-end counterparts, and while imitation transmits high resemblance, their perceived quality erodes. The perceived lower quality of clones could be due to the substitution of cheaper ingredients with poorer scent attributes, or alternatively to the influence of pricing and branding on hedonic judgment \cite{plassmann2008marketing,kim2026role}. Furthermore, imitative selection has weakened over time. During the complexity transition of 1985–2004, a strong negative association between note selection and prevalence suggests anti-conformist, novelty-seeking selection \cite{acerbi2014biases}, with saturated olfactive motifs becoming active targets of avoidance, and in response cultural exploration widens, as the rise in perfume surprisals implies. These dynamics place perfume alongside music, patents, and fashion as a cultural system in which novelty and conformity trade off \cite{gonzalez2023quantifying,barron2018individuals,park2023papers}.

Perfumery style may conform to the "three degrees of influence" rule, similar to phenomena previously documented in examples of social contagion \cite{christakis2007spread, christakis2008collective, fowler2008dynamic,fowler2010cooperative,airoldi2024induction}. Beyond the third degree of separation, perfumers are less likely to share stylistic similarities than random chance would predict, which mirrors patterns found in other systems \cite{christakis2007spread, christakis2008collective, fowler2008dynamic,fowler2010cooperative,airoldi2024induction}.

Our approach has limitations. First, our dataset lacks data on the discontinuation dates of the perfumes. The majority of perfumes and niche brands do not achieve commercial success, and this leads to the discontinuation of the products. In addition, regulatory restrictions force some perfumes to be discontinued. In both cases, discontinuation removes a lineage from the population, analogous to extinction in biology. Lack of this information prevents us from further analysis of cultural selection forces in perfume survival. For example, our measure of selection is based on a cultural-resemblance fitness rather than commercial survival. However, the commercial success of Original blockbusters usually attracts more imitation attempts to carve part of the market; thus, these two selection pictures should be intimately related, but without discontinuation data, it's not possible to make a stronger conclusion. Second, our dataset only includes marketing material on perfume notes and accords rather than exact perfume formulas. The exact perfume formulas are only available to perfume houses and brands. Although crowd-sourced marketing materials are mapping  perfume ingredients into a primary semantic scent descriptor space, we cannot confirm that all the findings of this work hold for formal perfume ingredients. Third, the resemblance annotations are crowd-sourced and incomplete, and because the platforms are contemporary, the resolution is not strictly comparable across eras. Fourth, for some perfumes only the release decade may have been recorded; the peaks at each decade's first year in Figure~\ref{fig1}A likely reflect dates defaulting to the decade's start.  Finally, the data are aggregated at a global level, which prevents access to local preferences. This limits our ability to examine fine-grained geospatial patterns, temporal trends, or socio-cultural influences \cite{majid_human_2021,arshamian_perception_2022,drnovsek2025demographic,henrich2010weirdest}.

Perfume and perfumery will continue to evolve. Similar to the effect of technology in driving perfumery from its classic form to industrial perfumery, recent advances in artificial intelligence (AI) may signal the start of a new evolutionary epoch in perfumery. In part, these advances build on the power of language in describing odor, demonstrated across the prediction of human olfactory perception \cite{keller2017predicting}, the principal odor map (POM) \cite{lee2023principal}, olfactory mixture-similarity prediction \cite{satarifard2025high}, generative perfume-design techniques \cite{sharma2025navigating,aleixandre2025generative,rodrigues2024molecule}, and the growing capabilities of large language models \cite{makri2026benchmark} in olfactory tasks. These advances may signal the dawn of what we might call the "Generative Perfumery Era."

\section{Methods}\label{Methods}
\subsection{Perfume Dataset}
We compiled perfume records from three online databases: Fragrantica (\url{https://www.fragrantica.com}), Parfumo (\url{https://www.parfumo.com}), and Aromo (\url{https://aromo.ru}). In addition, perfume ingredient list is obtained from the Fragrance Ingredient Glossary (FIG), April 2020 Edition (\url{https://ifrafragrance.org/}). A detailed comparison of the three databases is provided in the \hyperref[Supplementary information]{Supplementary information}. Main body of the analyses is based on Fragrantica, while Parfumo, Aromo, and FIG are used for validation and comparison.
\subsection{Perfume Complexity and Novelty}
To quantify perfume complexity and novelty, we represent each perfume by its constituent notes. We quantified a perfume's olfactory complexity as the Shannon entropy of
its note distribution,
\begin{equation*}
     H = - \sum_{i} p_i \log_2 p_i,
\end{equation*}
where $p_i$ is the weight of note $i$ normalized over all positive notes. Higher $H$ means a more complex note distribution.

We quantified novelty as how unexpected a perfume's note combinations are against all earlier formulations. Each note is normalized by total vote share $w_i$, and each pair $(i,j)$ is defined by the average vote weight $w_{ij}=(w_i+w_j)/2$. The surprisal is $s_{ij}=-\log_2 p_{ij}$, where $p_{ij}=(n_{ij}+1)/(n+2)$ is the smoothed frequency across the $n$ preceding perfumes and $n_{ij}$ is the number of those perfumes containing both notes. A perfume's novelty is defined as the vote-weighted mean surprisal over all its pairs
\begin{equation*}
\mathrm{Novelty} = \frac{\sum_{ij} w_{ij}\, s_{ij}}{\sum_{ij} w_{ij}},
\end{equation*}
so rarer combinations score higher.  We computed the analogous triple measure using 
$w_{ijk}=(w_i+w_j+w_k)/3$. We calculated both complexity and novelty for highly-voted perfumes to reduce the noise (i.e., $\mathrm{N_{votes}} \geq 286$, the average number of votes per perfume from Fragrantica). An unweighted variant and a Word2Vec-based measure of novelty \cite{mikolov2013efficient} are given in the \hyperref[Supplementary information]{Supplementary information}.
\subsection{Lineage Network of Perfume.}
We treated each resemblance annotation (i.e., "reminds-me-of" annotation in Fragrantica) as a directed copy edge from the older perfume (the Original) to the newer one (the flanker or clone). We discarded pairs without a valid release-year information. We restricted edges to highly-voted perfumes. For any given two perfumes we measured note overlap by the Jaccard index, and kept a copy edge only when $J \geq J^{\ast}$, where $J^{\ast}=0.39$ is the 95th percentile  of $J$ over all resembled pairs. Each perfume was assigned its highest Jaccard index qualifying predecessor as its primary parent and, where present, the next-best qualifying predecessor as a secondary parent. We treated perfumes rated as great value for their price-value as lineage terminals and not used as parents, this was taken as a sign of a low-cost clone.
\subsection{Selection Differential}
We quantified copy-selection on individual notes with the Price-equation selection differential~\cite{price1970selection}. Within each era we scored each perfume $i$ by its cultural fitness $w_i$ (i.e., the number of later perfumes in that era that cite it via a resemblance annotation) and set $z_{ik}=1$ if $i$ carries note $k$. The selection differential of note $k$ is its fitness-weighted minus unweighted mean presence,
\begin{equation*}
  S_k = \frac{\sum_i w_i z_{ik}}{\sum_i w_i} - \frac{1}{N}\sum_i z_{ik}
      = \frac{\mathrm{cov}_i(w_i, z_{ik})}{\bar{w}},
\end{equation*}
where $N$ is the count of all perfumes in that era, $S_k>0$ indicates over-copied notes and $S_k<0$ under-copied ones, we scored only notes carried by at least ten perfumes in the era. Significance was assessed against an analytic fitness-permutation null~\cite{cochran1977sampling}, with $p$-values FDR-corrected by Benjamini-Hochberg. We computed each note in-era trend as the change in the LOWESS-smoothed prevalence between the first and last year, divided by the span ($\Delta$ prevalence per year).

\subsection{Consolidation-Disruption (CD) Index}
We treated time-directed resemblance annotation as a citation. We then applied the CD index~\cite{park2023papers} to the network of highly-voted perfumes linked at note overlap of the 90th percentile $J^{\ast}=0.32$. We used a lower threshold here than for the lineage network, since the CD index requires a denser citation structure to be estimated reliably. For a focal perfume $F$ and forward window $t$, we define $f_i=1$ if a later perfume cites $F$ and $b_i=1$ if it cites a predecessor of $F$, the CD is calculated by $\mathrm{CD}_t = (1/n_t)\sum_i \left( f_i - 2 f_i b_i\right)$, where $n_t$ counts perfumes citing $F$ or a predecessor. Positive and negative CD indicates disruption and consolidation, respectively.

\subsection{Accord and Note Contagion Across Social Distance}
We summarized each perfumer's style by their five most frequent notes and accords. To separate social transmission from pre-existing similarity, we used a directional, event-anchored test with a matched null. For each ordered pair at social distance $d$, we defined focal perfumer $A$ and source perfumer $B$, connected in a given year by a collaboration chain with years non-decreasing toward $A$ (forward in time only). We asked whether any of $B$'s pre-collaboration top-five descriptors absent from $A$'s enters $A$'s post-collaboration top five (excluding joint perfumes and requiring at least two perfumes for $A$'s post-collaboration and for $B$'s pre-collaboration). The adoption probability $P_{\mathrm{obs}}(r,d)$ is the fraction of such rank-$r$ descriptors that subsequently enter $A$'s top five. In the matched null, $B$ is replaced by a random non-collaborator of $A$ active before the collaboration year, matched on shared pre-collaboration top-five descriptors and shared brand affiliation, giving $P_{\mathrm{rand}}(r,d)$. A positive residual reflects contagion beyond homophily and brand affiliation. We report $\mathrm{Adoption~Probability}(r,d) =\big((P_{\mathrm{obs}} - P_{\mathrm{rand}})/P_{\mathrm{rand}}\big)\times 100$. We ran $n = 5000$ simulations, considering changes significant when $|z| \geq 1.96$, where $z = (P_{\mathrm{obs}} - P_{\mathrm{rand}})/\sigma_{\mathrm{rand}}$.

\subsection{Standardised OLS Effect-Size Models}
Effect sizes were estimated with ordinary least-squares models in which all variables were z-scored before estimation. For predictors of perfume popularity we fit 
\[
\text{rating}\sim\text{longevity}+\text{sillage}+\text{price}+\text{complexity}+\text{novelty}.
\]
For drivers of collaboration we fit:
\[
\text{collaboration}\sim\text{brand}+\text{country}+\text{era}+\text{note}+\text{accord},
\]
where brand is the perfumer's brand affiliation, country is their most listed country (modal), era is the negative gap between their median release years, and note and accord similarity are cosine similarities of their note and accord frequency vectors. For each model we report standardised coefficients with $95\%$ confidence intervals and $p$-values are FDR-corrected by Benjamini-Hochberg.

\bmhead{Supplementary information}\label{Supplementary information} The supplementary material is available on pages~S1-S19.

\bmhead{Acknowledgements} V.S. thank Alex Wiltschko for very helpful feedback on the manuscript.

\section*{Declarations}

%Some journals require declarations to be submitted in a standardised format. Please check the Instructions for Authors of the journal to which you are submitting to see if you need to complete this section. If yes, your manuscript must contain the following sections under the heading `Declarations':

\begin{itemize}
\item \textbf{Funding:} This research was supported in part by grants from  the NOMIS Foundation, Pershing Square Philanthropies, the Stavros Niarchos Foundation, Rothberg Catalyzer, Paul Graham Foundation, and Schmidt Futures. 
\item \textbf{Conflict of interest/Competing interests:} L.S. is currently the CTO of the startup Patina; she was not affiliated with Patina when the initial work on this manuscript was conducted. C.L. is a master perfumer at DreamAir and Osmo Labs.
%\item Ethics approval and consent to participate
%\item Consent for publication
\item \textbf{Data  and Code availability:} Partially anonymized datasets, models, evaluation scripts to generate all figures are available at the following repository \url{https://github.com/Satarifard/Cultural-Evolution-of-Perfumes}
\item \textbf{Author contribution:} V.S and F.B conceived the idea and designed the study. V.S. and L.S. collected and organized the datasets. V.S. conducted the analysis and generated the figures. All authors interpreted the results of the study. V.S., F.B., L.M.H, and G.M. wrote the paper with input and revisions from all the authors.  C.L. and N.A.C. supervised the research.
\end{itemize}

\bibliography{sn-bibliography}

%%%%%%%%%%%%%%%% END OF MAIN TEXT %%%%%%%%%%%%%%%

\newpage

%%%%%%%%%%%%%%%% START OF SUPPLEMENT %%%%%%%%%%%%%%%
\renewcommand{\thefigure}{S\arabic{figure}}
\renewcommand{\theHfigure}{S\arabic{figure}}   
\renewcommand{\thetable}{S\arabic{table}}
\renewcommand{\theHtable}{S\arabic{table}}
\renewcommand{\theequation}{S\arabic{equation}}
\renewcommand{\theHequation}{S\arabic{equation}}
\renewcommand{\thepage}{S\arabic{page}}
\renewcommand{\thesection}{S\arabic{section}}
\setcounter{section}{0}
\setcounter{figure}{0}
\setcounter{table}{0}
\setcounter{equation}{0}
\setcounter{page}{1}

%%%%%%%%%%%%%%%% SUPPLEMENT TITLE PAGE %%%%%%%%%%%%%%%

\begin{center}
\section*{Supplementary Information for\\ \scititle}

Vahid Satarifard$^{\ast}$,
Fabian Baumann, 
Geetanjali Minsky,\\
Laura Sisson, 
Lou M. Haux, 
Christophe Laudamiel, and
Nicholas A. Christakis,\\

\small$^\ast$Corresponding author. Email: vahid.satarifard@yale.edu\\
\end{center}

\section{Dataset comparison}
Several publicly available online perfume datasets exist, which are primarily designed as perfume encyclopedias. Among those the most reliable datasets are Fragrantica (\url{https://www.fragrantica.com}), Parfumo (\url{https://www.parfumo.com}), and Aromo (\url{https://aromo.ru}). Another highly regarded database, Fragrances of the World (\url{https://www.fragrancesoftheworld.info/}), curated by Michael Edwards, is not publicly available. We contacted Edwards about using this dataset, but he declined to share it due to the risk of data leakage. We therefore assembled perfume records from the three publicly available databases, i.e., Fragrantica, Parfumo, and Aromo. To test whether the annotations in these datasets reflect a genuine signal rather than the conventions of a single community, we compared them against one another. The comparison across all three databases is summarized in Table \ref{tab:dataset_comparison}. Before computing any cross-database quantity, we mapped all notes and accords to a common canonical vocabulary to resolve differences in transliteration, pluralization, and spelling, as well as known perfumery pseudonyms. This mapping is not complete or bias-free, however, given the large size of the datasets.

\begin{figure}[!htbp]
\centering
\includegraphics[width=1\textwidth]{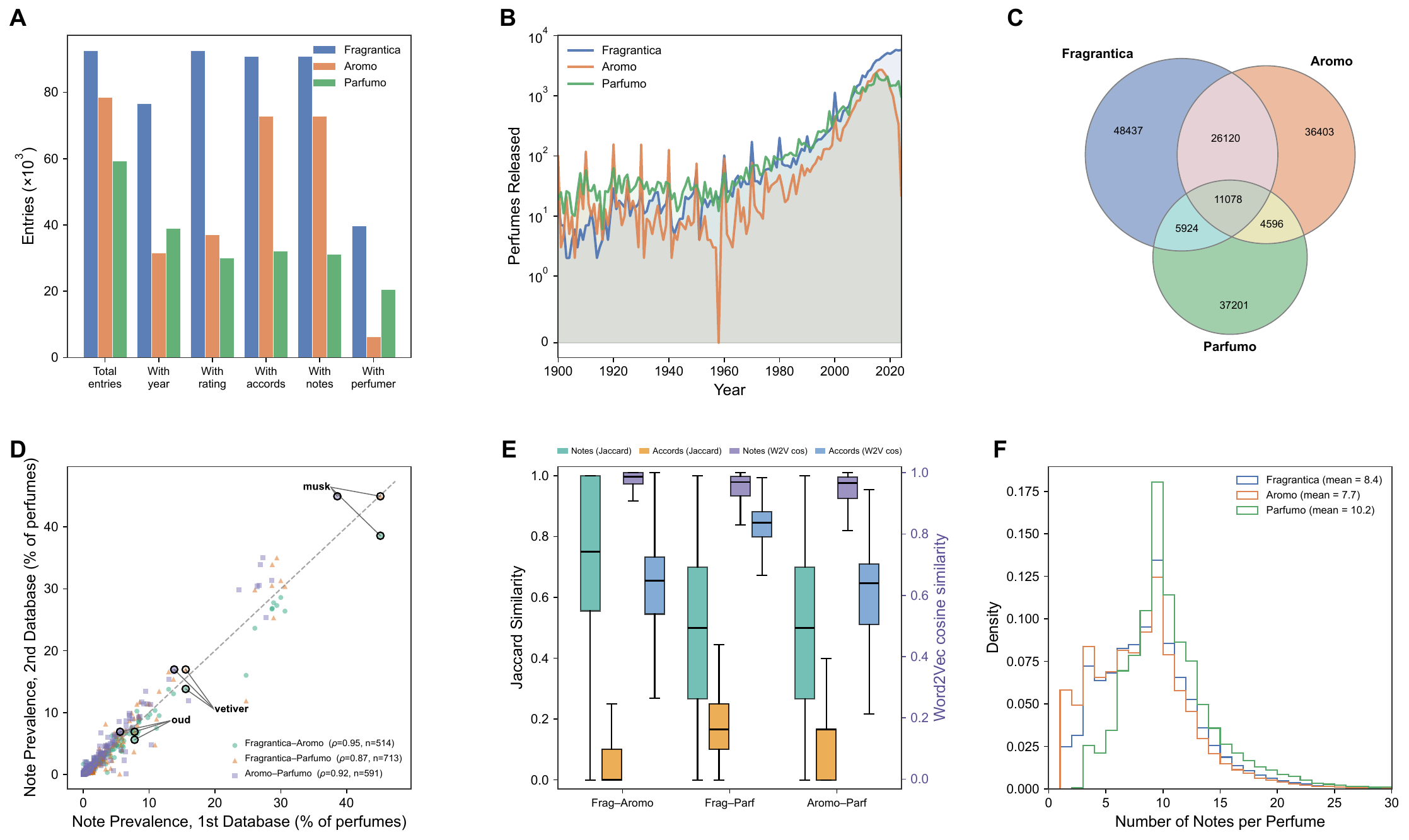}
\caption{\textbf{Cross-database comparison of the three perfume catalogues.} \textbf{(A)} Field coverage of Fragrantica (n = 92{,}589), Aromo (n = 78{,}409), and Parfumo (n = 59{,}280) (entries with a release year, rating, accords, notes, perfumer). \textbf{(B)} Perfumes released per year, 1900--2024  \textbf{(C)} Catalogue overlap from exact (brand, perfume name) matching with 11,078 perfumes common to all three, and 37,198 for Fragrantica-Aromo, 17,002 for Fragrantica-Parfumo, and 15,674 for Aromo-Parfumo) common perfumes, respectively. \textbf{(D)} Shared-note prevalence in one database versus the other for each pair. \textbf{(E)} Agreement between matched perfumes' note and accord sets with exact Jaccard index (left axis; median note 0.75 / 0.50 / 0.50, accord 0.00 / 0.17 / 0.17) and Word2Vec centroid cosine similarity (right axis; median note 0.97--0.99, accord 0.64--0.84). \textbf{(F)} Area-normalised histograms of notes per perfume.}\label{fig:S1}
\end{figure}

The three datasets differ in size and in the breadth of metadata they provide (Table \ref{tab:dataset_comparison} and Figure \ref{fig:S1}A). Fragrantica is the most comprehensive, comprising 92,589 perfumes with more than 98\% of perfumes covering perfume notes and accords, and user ratings, together with other metadata which are absent from the other two sources, such as gender classification, country of origin, and market segment (Table \ref{tab:dataset_comparison}). It is also by far the most heavily annotated by its community, aggregate crowdsourced totals exceed 16 million overall rating votes, and 25 million votes on perfume notes, alongside votes on sillage, longevity, price value, and gender association. In addition, Fragrantica includes resemblance annotations, which is a community judgment that one perfume reminds the rater of another, which we use to construct copy lineage networks. Parfumo and Aromo are smaller,  with 59,280 and 78,409 perfume records, respectively. They also provide perfume pyramids, ratings, and family/accord tags, but lack crowd-sourced perceptual data.

The three catalogues are similar on the descriptors central to our analyses. Annual release counts follow a common, near-exponential trajectory across the twentieth century (Figure \ref{fig:S1}B). The Aromo data is noisier, and the sharp drop in the number of entries between 2010 and 2024 indicates incomplete records for that period. In addition, the peaks in Figure \ref{fig:S1}B suggest that, for a portion of perfumes, only the release decade was known, this is why entries peak on the first year of each decade. The exact matching on brand and perfume name allows us to recover a large shared core of 11,078 perfumes common to all three databases (Figure \ref{fig:S1}C). The prevalence of individual notes is strongly correlated, with Spearman $\rho$ ranging 0.87–0.95, between every pair of databases (Figure \ref{fig:S1}D).  For perfumes present in two databases, the exact overlap of note sets is substantial. The median note Jaccard ranges from 0.50 for Fragrantica-Parfumo and Aromo-Parfumo to 0.75 for Fragrantica-Aromo (Figure~\ref{fig:S1}E). Exact agreement on accord labels is much weaker, with median accord Jaccard between 0.00 (Fragrantica-Aromo) and 0.17. However, this gap seems to be largely due to difference in vocabulary usage, as we did not enforce a fully harmonised canonical vocabulary across platforms, so synonyms, spelling variants, and labels at different levels of granularity are scored as mismatches by exact-string Jaccard. To test this idea, we used an embedding of each perfume's notes and accords with Word2Vec \cite{mikolov2013efficient} and compared the pair of database's by cosine similarity of their embeddings. The note descriptions are nearly identical in every pair with median cosine ranging 0.97 to 0.99, and accord agreement is much higher than Jaccard implies, with median cosine ranging 0.64 to 0.84. Aromo is a Russian-language source, and the accord cosine similarity is $\approx$0.64 for both Aromo pairs versus 0.84 for Fragrantica-Parfumo. This suggests that translation and taxonomy differences (Aromo uses family instead of accord) lead to this lower score rather than genuine disagreement about a perfume's underlying scent perception. Finally, all three datasets show a comparable granularity with 7.7–10.2 notes per perfume (Figure \ref{fig:S1}F). 

These comparisons show that Fragrantica is the most suitable dataset for main-body of analysis in our study. Thanks to Fragrantica's larger user base and international traffic it provides the most richly annotated of the publicly accessible perfume databases. The close agreement of two independently compiled databases on release timing, catalogue composition, and note and accord demonstrates that Fragrantica's annotations capture a reproducible signal rather than the unique features of a single community. We therefore base the main analysis on Fragrantica, and when possible, we use Parfumo and Aromo for validation.

\section{Estimation of carrying capacity of perfume industry}
\begin{figure}[!htbp]
\centering
\includegraphics[width=1\textwidth]{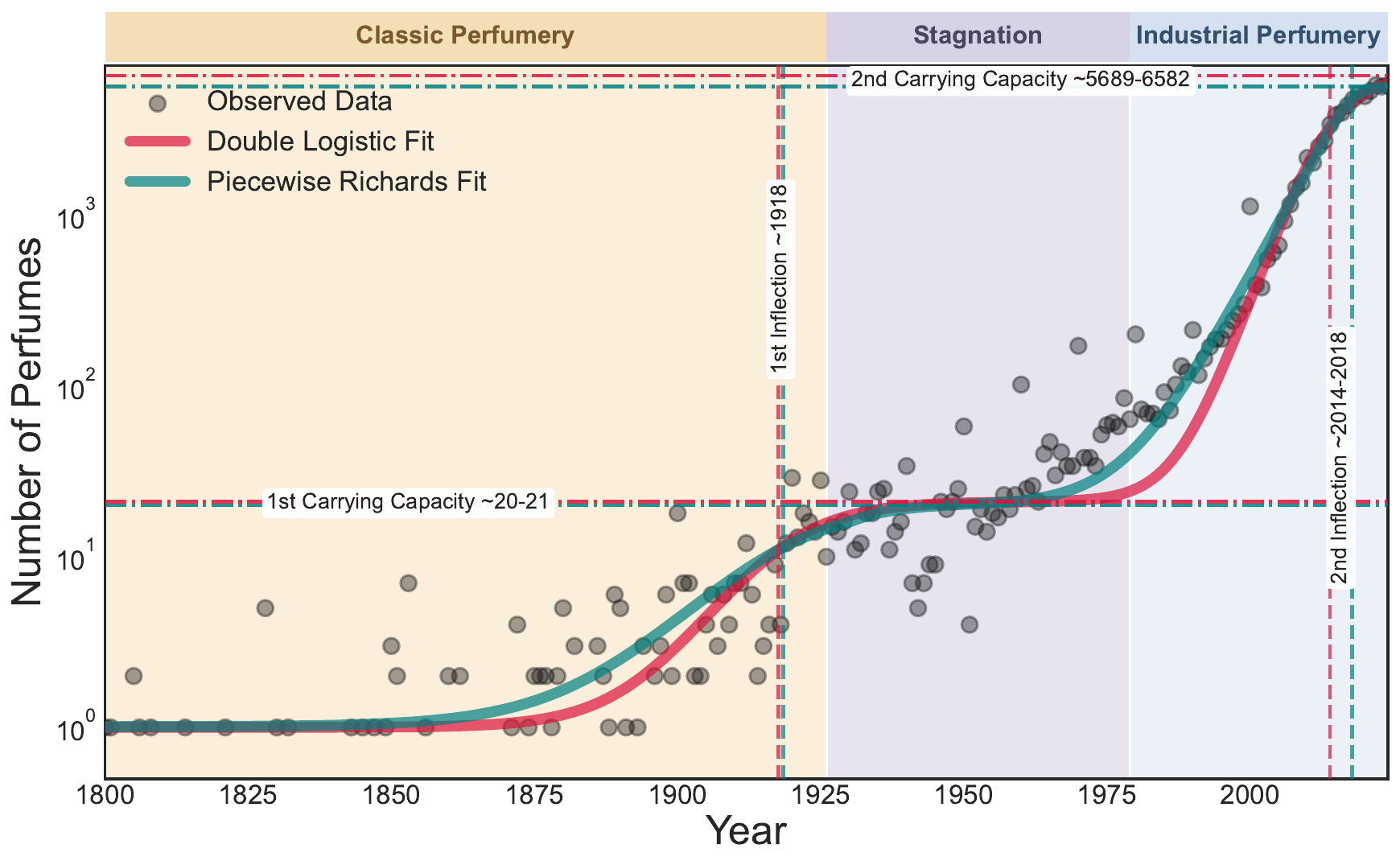}
\caption{\textbf{Perfume Carrying Capacity.} Yearly perfume releases from 1800 to 2024 (dark gray points). The red and green solid lines show the double-logistic and piecewise-Richards model fits, respectively. Background shading partitions the record into three eras, Classic Perfumery era (before 1926), Stagnation period (1926 to 1979), and Industrial Perfumery era (after 1979). The stagnation window  is defined directly from the data as the continuous interval between the two inflection points where the fitted log-growth rate falls below 15\% of its peak. Vertical dashed lines indicate the inflection points, and horizontal dash-dotted lines indicate the carrying capacities, each colored to match its model. The first inflection occurs around 1918, indicating a rise to an initial carrying capacity of roughly 20 to 21 perfumes per year. A second inflection appears at 2014 (double logistic) and 2018 (piecewise Richards), with carrying capacities of approximately 6,582 and 5,689 perfumes per year, respectively.}\label{fig:S2}
\end{figure}

We characterize the long-term trends of perfume production in greater detail by fitting data to both a double logistic model and a piecewise Richards model \cite{richards1959flexible} for annual number of perfumes introduced between 1800 and 2024 (Figure~\ref{fig:S2}). These methods allow us to identify critical macroscopic features of the dynamics of perfume production, in particular, inflection points, growth phases, and carrying capacities. 

For double-logistic model, we modelled the annual totals \(N(t)\) with a superposition of two logistic growth curves sharing a free lower asymptote \(A\):
\[
N(t)=A+\frac{K_{1}}{1+\exp\!\bigl[-r_{1}(t-t_{1})\bigr]}+\frac{K_{2}}{1+\exp\!\bigl[-r_{2}(t-t_{2})\bigr]}
\]
Here \(K_{j}\) is the carrying capacity, \(r_{j}\) the intrinsic growth rate, and \(t_{j}\) the inflection year of wave \(j\in\{1,2\}\).
Parameter estimation employed non-linear least squares from SciPy \cite{virtanen2020scipy}. In addition, because the post-1950 trajectory departs from pure logistic behavior, a pair of generalized logistic (Richards) functions was fitted to two overlapping windows, namely Wave 1 (1800–1950) and Wave 2 (1930–2024).  
The Piecewise Richards reads \cite{richards1959flexible}
\[
N(t)=A+\frac{K-A}{\bigl[1+\exp\!\bigl(-r\,(t-t_{0})\bigr)\bigr]^{\nu}}
\]
where  \(A\) is the lower asymptote,  
\(K\) is the upper (carrying-capacity) asymptote, \(r\) the intrinsic growth rate,  \(t_{0}\) is the inflection year, and \(\nu\) is the shape parameter which governs curve asymmetry. The two fits were combined into a single continuous curve by linear blending between 1930 and 1950. Both models were evaluated on a dense grid of 3,000 points spanning 1800–2024. 

The model fits reveal two distinct plateaus in the production dynamics and the associated inflection points (Figure~\ref{fig:S2}). The first plateau spans from approximately 1926 to 1979, a period of stagnation with an average of about 20 perfumes per year following an earlier rise with an inflection point around 1918. The second inflection point occurs between 2014 and 2018, identifying the onset of a rapid expansion that levels off at a new carrying capacity of roughly 6,000 perfumes per year. These dynamics point to three major historical phases in the evolution of commercial perfumery: (i) the Classic Perfumery Era (before 1926), characterized by gradual growth, (ii) the Stagnation Period (1926 to 1979), identified by limited innovation and production, and (iii) the Industrial Perfumery Era (after 1979), defined by a surge in output, the emergence of large-scale fragrance manufacturing, and consolidation in the industry through mergers or acquisitions \cite{jones2010beauty} (Figure~\ref{fig:S2}). This classification is obtained from our analysis and is used here for simplicity, these are not commonly accepted terms among perfume historians.

\section{Regional profiles of scent}

\begin{figure}[!htbp]
\centering
\includegraphics[width=1\textwidth]{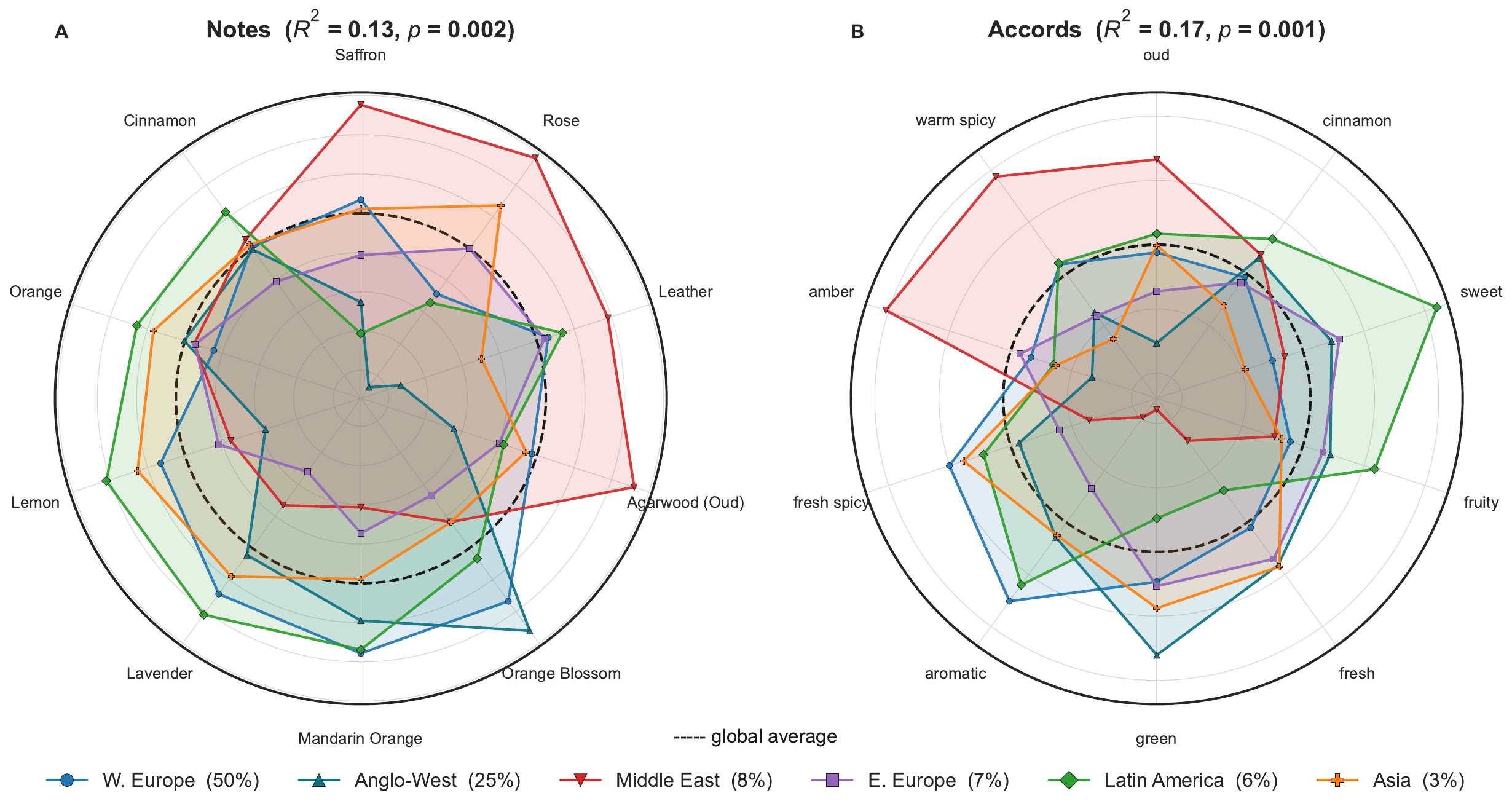}
\caption{\textbf{Regional profiles of scent.} The six regions of Figure 1D are each summarized by their mean profile and overlaid on one radar, \textbf{(A)} for notes and \textbf{(B)} for accords. Values are z-scored log location quotients (a region's shares relative to the global average), so the dashed ring is the global mean and points beyond it indicate over-use of that note or accord and vice versa. The ten axes per panel are the highest-loading items in the principal-component space, arranged by their direction there so similar items sit together (e.g. oud, saffron, leather and cinnamon on one side, citrus and fresh items on the other). Regions follow the colour and marker scheme of Figure 1D. The Middle East over-indexes on oriental materials (oud, saffron, leather, rose; oud, amber, warm spicy), Latin America on sweet and fruity, and the Anglo-West on fresh and green. By PERMANOVA, region explains 13\% of the variance in note profiles ($p = 0.002$) and 17\% in accord profiles ($p < 0.001$).}\label{fig:S3}
\end{figure}

\begin{figure}[!htbp]
\centering
\includegraphics[width=1\textwidth]{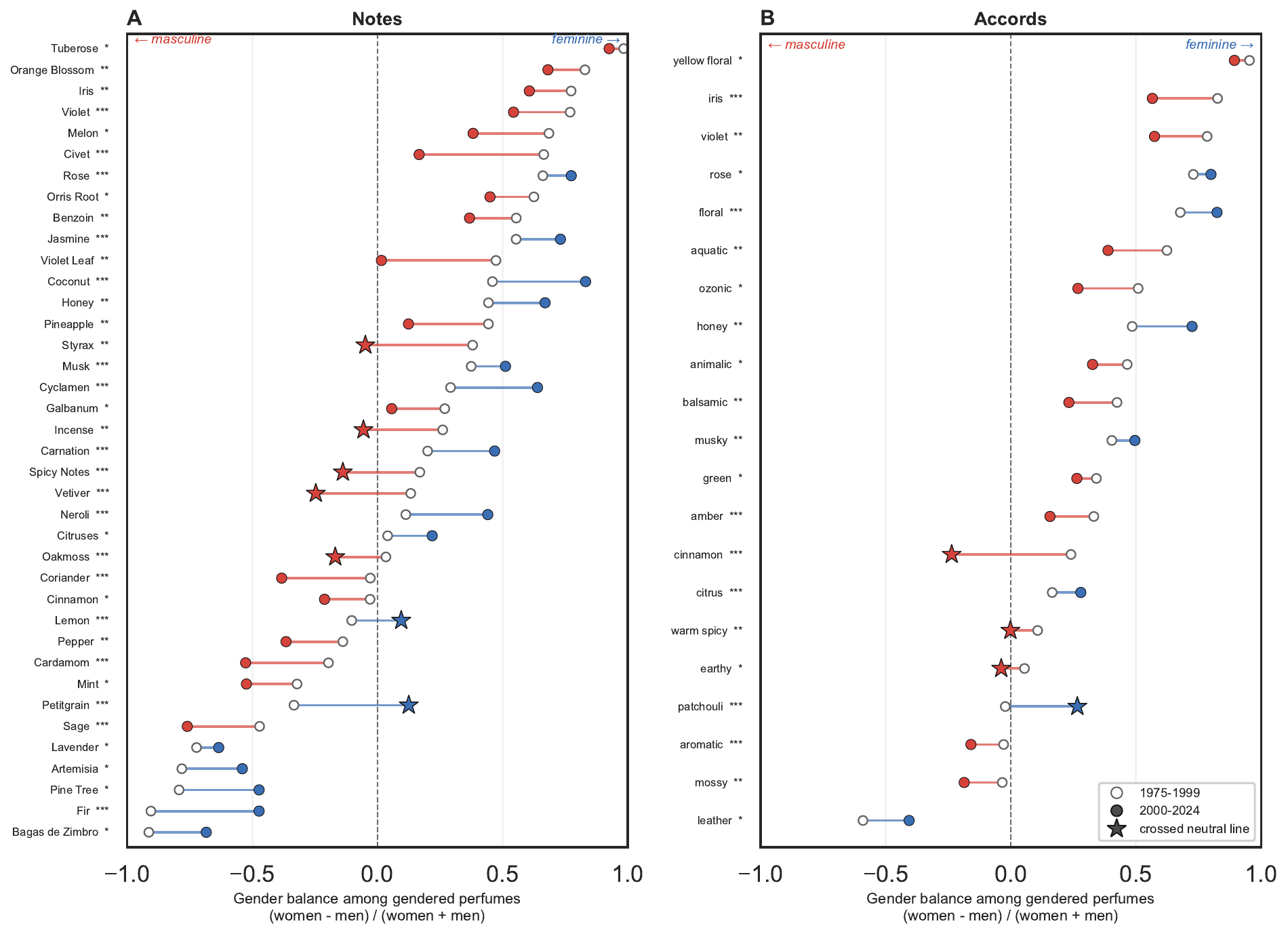}
\caption{\textbf{Gender coding of notes and accords.} For each note and accord we compute its gender balance among gendered perfumes, (women - men) / (women + men). Balances are compared between an early window (1975–1999, open circles) and a recent one (2000–2024, filled symbols). The connecting segment is red for shifts toward men and blue toward women. \textbf{(A)} Notes and \textbf{(B)} accords, ordered most masculine (bottom) to most feminine (top). Significance is per-item 2×2 chi-square (women vs men) across the two eras, significance is FDR corrected by Benjamini–Hochberg. Stars represent those crossing the neutral line. Descriptors that cross the neutral line include styrax, incense, spicy notes, vetiver, oakmoss, lemon, and petitgrain for notes, and cinnamon, warm spicy, earthy, and patchouli for accords}\label{fig:S4}
\end{figure}

\begin{figure}[!htbp]
\centering
\includegraphics[width=0.8\textwidth]{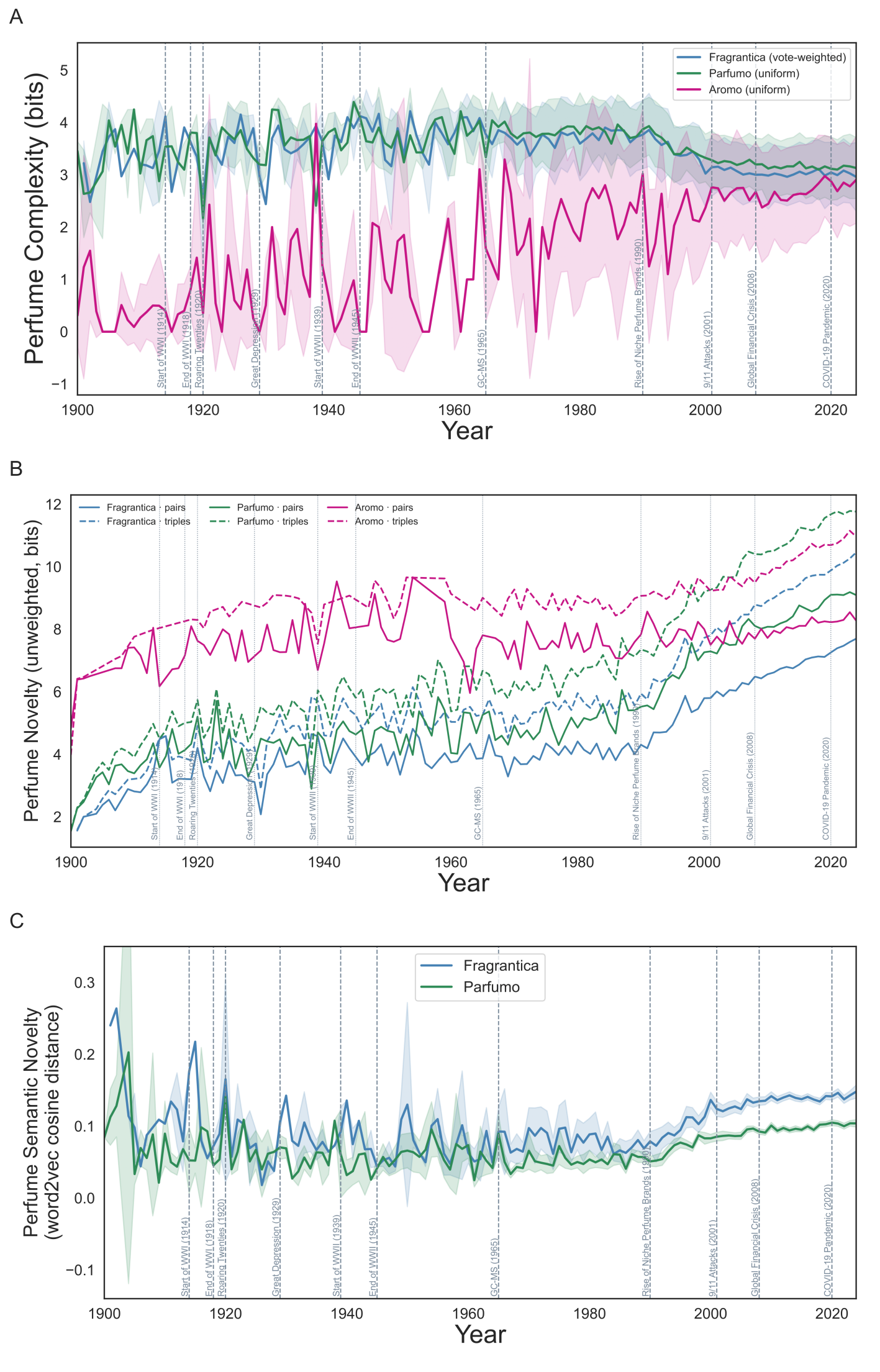}
\caption{\textbf{Cross-database robustness of the complexity and novelty trends.} Annual means of \textbf{(A)} complexity, \textbf{(B)} unweighted note co-occurrence novelty (note pairs solid, triples dashed), and \textbf{(C)} Word2Vec semantic novelty, computed for Fragrantica,  Parfumo and Aromo catalogues over 1900–2024. Shaded bands show standard deviation in \textbf{(A)} and 95\% confidence intervals in \textbf{(C)}. In \textbf{(A)}, Fragrantica uses vote-weighted entropy while Parfumo and Aromo, lacking per-note weights, use uniform weights where entropy equals log\textsubscript{2} of the note count. In \textbf{(C)}, Aromo is omitted because its largely non-English note labels embed poorly.}\label{fig:S5}
\end{figure}

\begin{figure}[!htbp]
\centering
\includegraphics[width=0.8\textwidth]{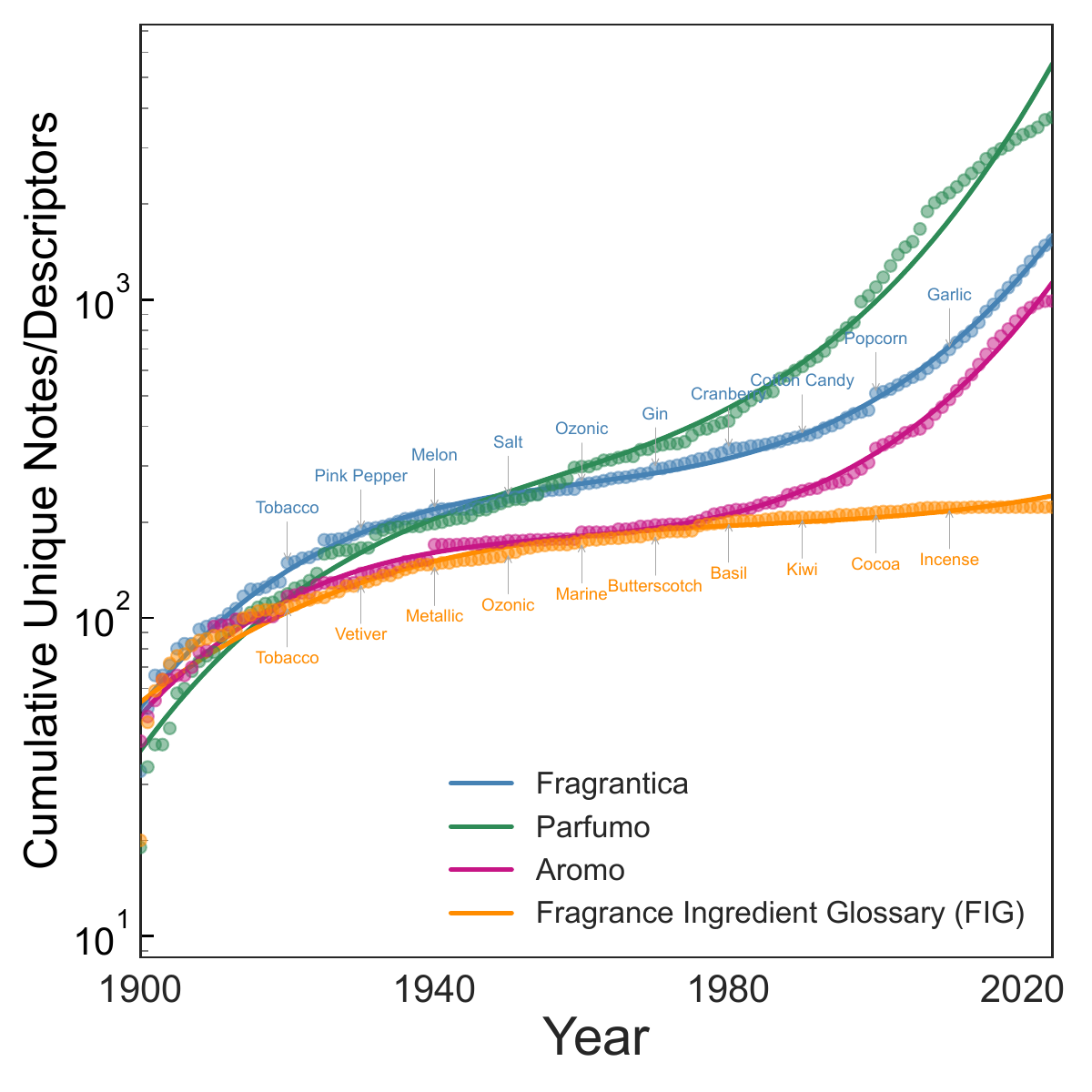}
\caption{\textbf{Growth of the olfactory palette.} Cumulative number of distinct notes across three perfume databases (i.e., Fragrantica, Parfumo, and  Aromo) and of distinct odor descriptors in the Fragrance Ingredient Glossary (FIG), plotted by each term's year of first appearance over 1900–2024. Points are observed cumulative counts and curves are degree-3 log–log fits. Labels indicate representative notes and descriptors introduced in successive decades in Fragrantica and FIG, respectively.}\label{fig:S6}
\end{figure}

\begin{figure}[!htbp]
\centering
\includegraphics[width=0.8\textwidth]{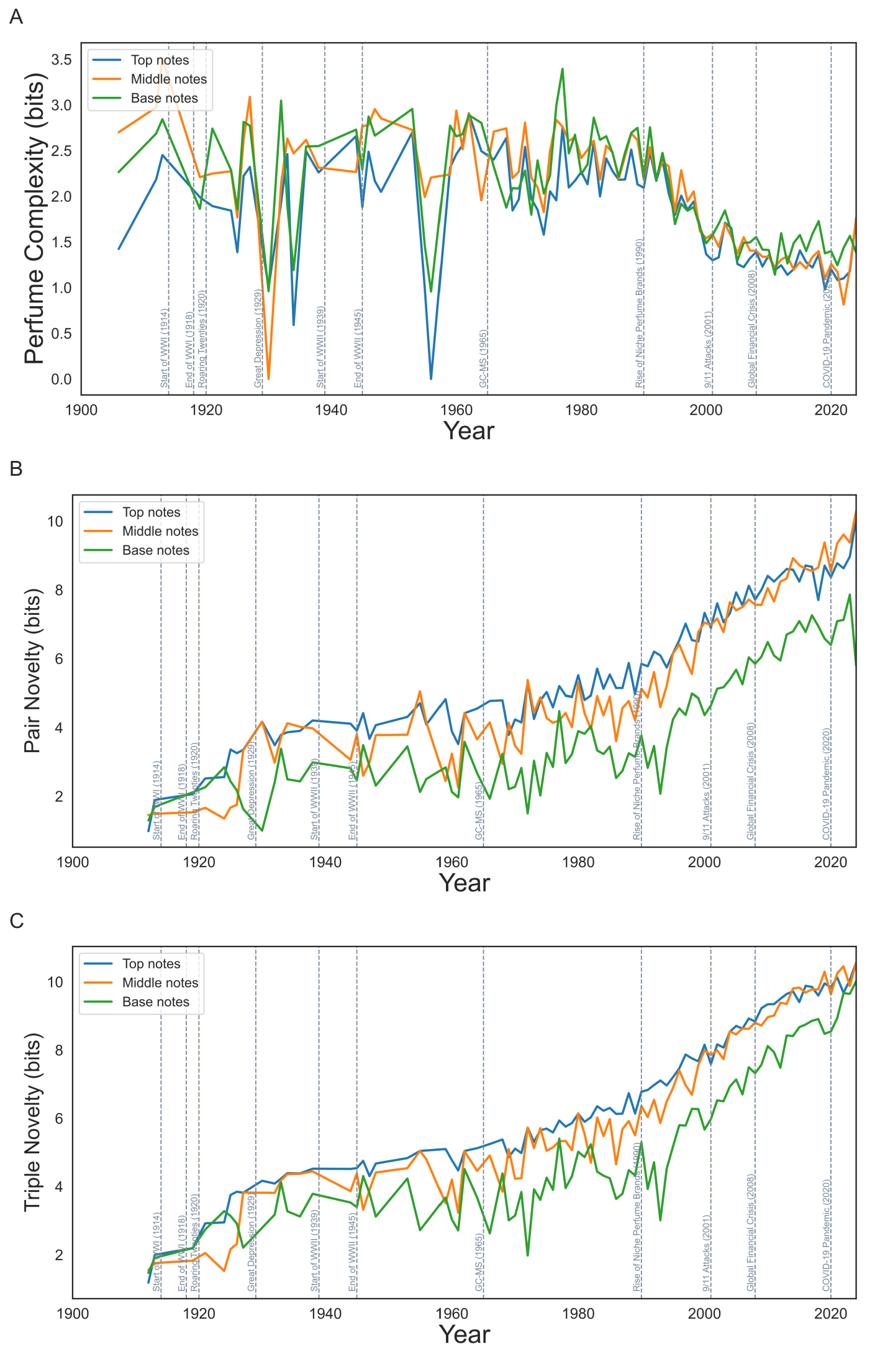}
\caption{\textbf{Complexity and novelty of perfume pyramid} Annual mean \textbf{(A)} complexity (Shannon entropy), \textbf{(B)} pair novelty, and \textbf{(C) }triple novelty, computed separately within the top, middle, and base layers of the perfume pyramid.}\label{fig:S7}
\end{figure}

\begin{figure}[!htbp]
\centering
\includegraphics[width=0.8\textwidth]{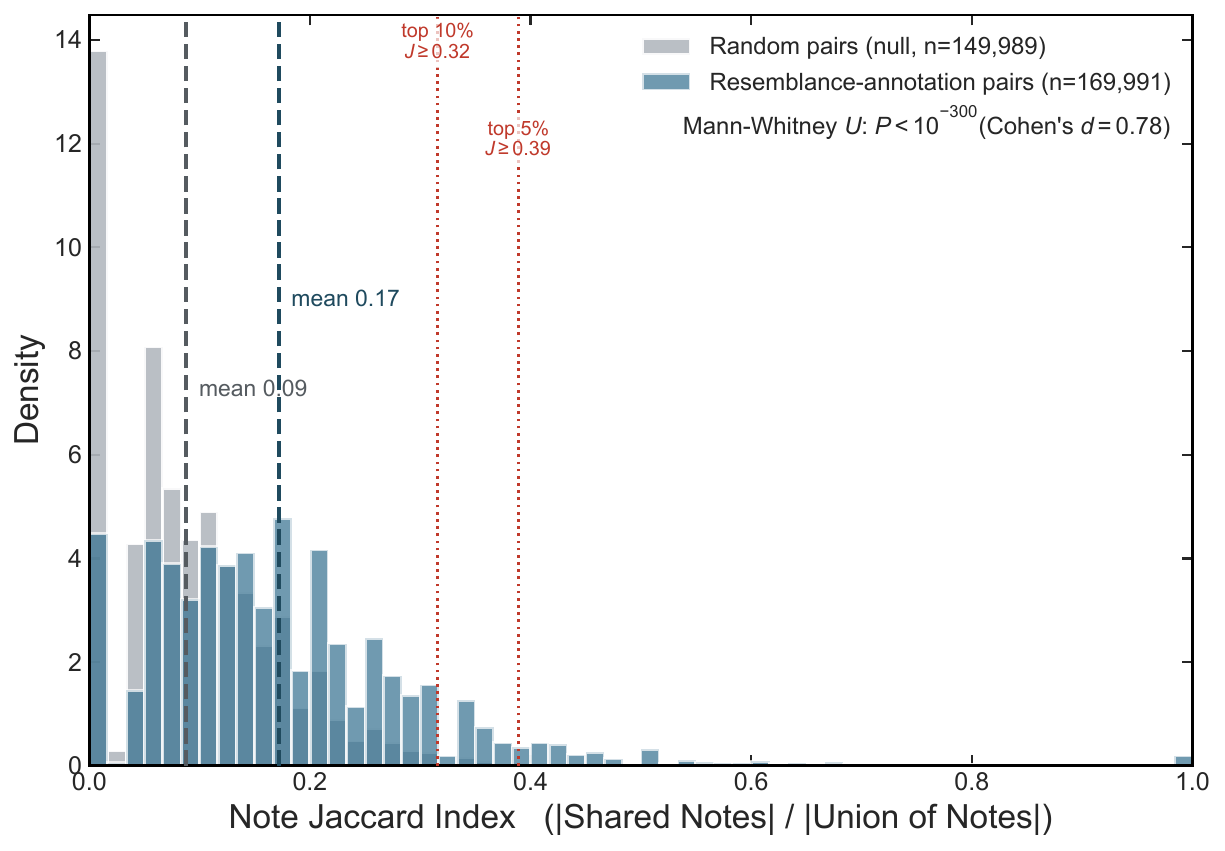}
\caption{\textbf{Note Jaccard Index distribution for random pairs and resembling perfumes.} (A) Distribution of the note Jaccard index for every resemblance-annotation pair among highly-voted perfumes ($\geq 286$ votes; $n = 169{,}991$, blue) versus a random-pair null from the same perfumes ($n = 149{,}989$, gray). Annotated pairs share far more notes than chance (means $0.17$ vs $0.09$; two-sided Mann--Whitney $U$, $P < 10^{-300}$, Cohen's $d = 0.78$). Red dotted lines show Jaccard index for top $10\%$ ($J \geq 0.32$) and top $5\%$ ($J \geq 0.39$). We use $J \geq 0.39$ for constructing the copy lineage networks in Figure~3A, and Figure \ref{fig:S9}.}\label{fig:S8}
\end{figure}

\begin{figure}[!htbp]
\centering
\includegraphics[width=0.8\textwidth]{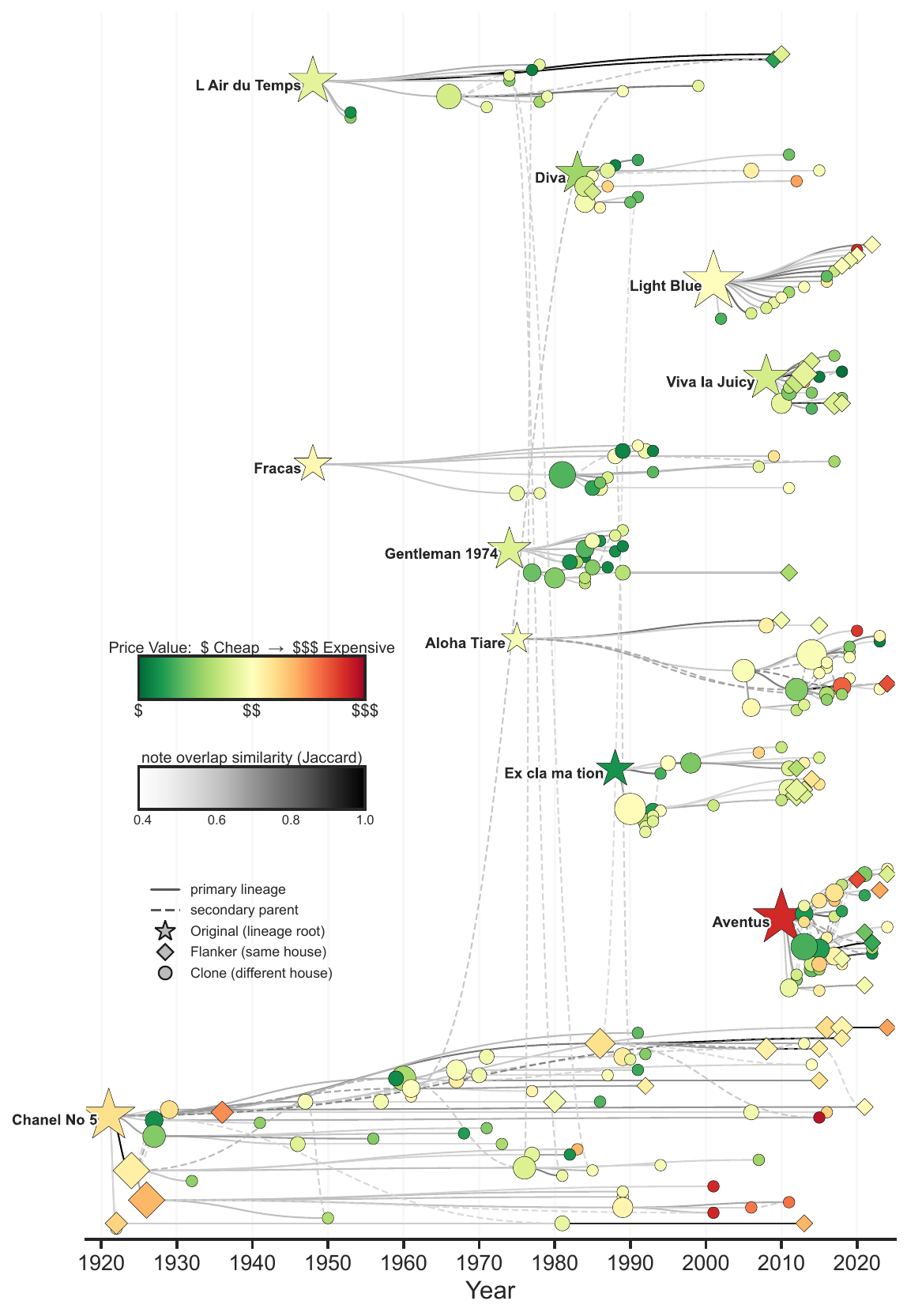}
\caption{\textbf{The top ten largest copy lineage networks.} 
This plot extends Figure 3A from the two copy lineage networks to the ten largest. The copy lineage networks are shown on a release-year axis. Solid curves link each perfume to its primary parent, dashed to a secondary parent. Symbol shape denotes Original (star), Flanker (diamond) or Clone (circle), and the size corresponds to the number of imitators, fill colour shows the price value (green to red denotes transition from great value to expensive) and edge shade the note overlap calculated from Jaccard index.}\label{fig:S9}
\end{figure}

\begin{figure}[!htbp]
\centering
\includegraphics[width=0.8\textwidth]{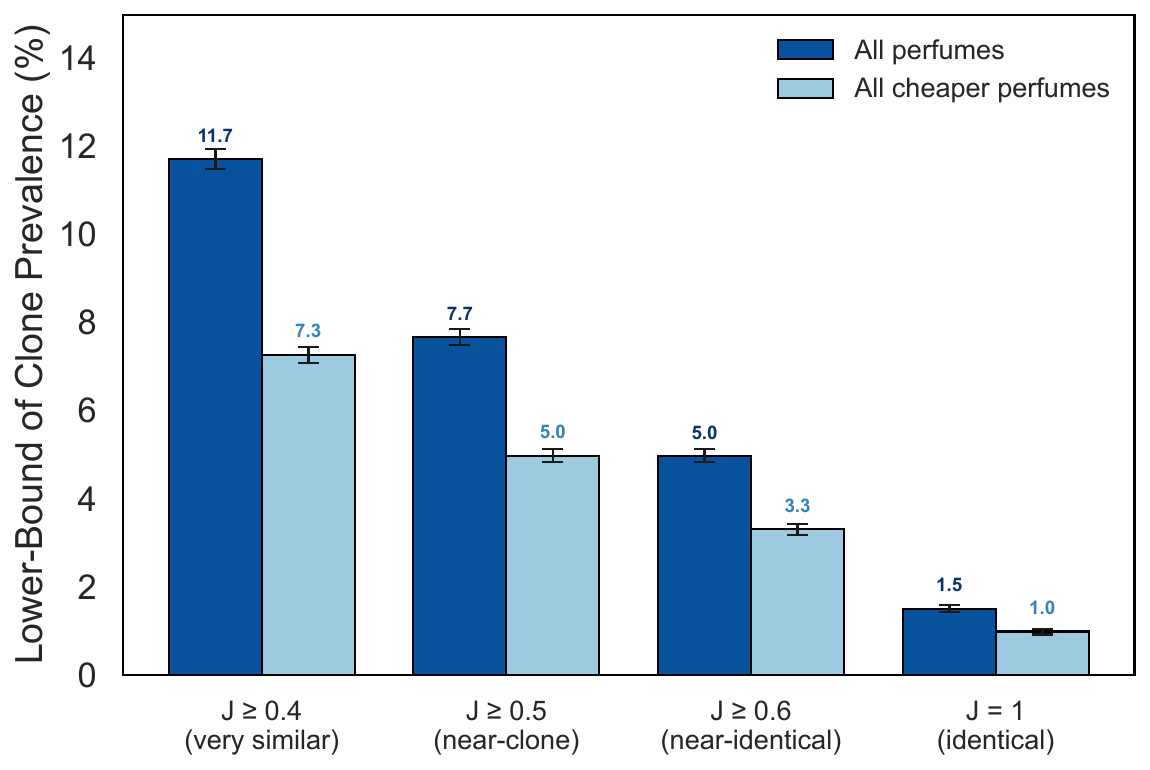}
\caption{\textbf{A lower bound for fraction of cloned perfumes.} Among the 75{,}673 perfumes with a note list and release year, a clone is one whose resemblance annotation points to an older, best-note-overlap perfume from a different brand (same-brand Flankers excluded). Bars give the clone share under progressively stricter note-overlap gates ($J \ge 0.4, 0.5, 0.6, 1$). All cross-brand clones: 11.7\%, 7.7\%, 5.0\%, 1.5\%. Restricting to clones rated higher in price value than their model: 7.3\%, 5.0\%, 3.3\%, 1.0\%. Error bars are bootstrap 95\% CIs. These are lower-bound estimates.}\label{fig:S10}
\end{figure}

\begin{figure}[!htbp]
\centering
\includegraphics[width=0.8\textwidth]{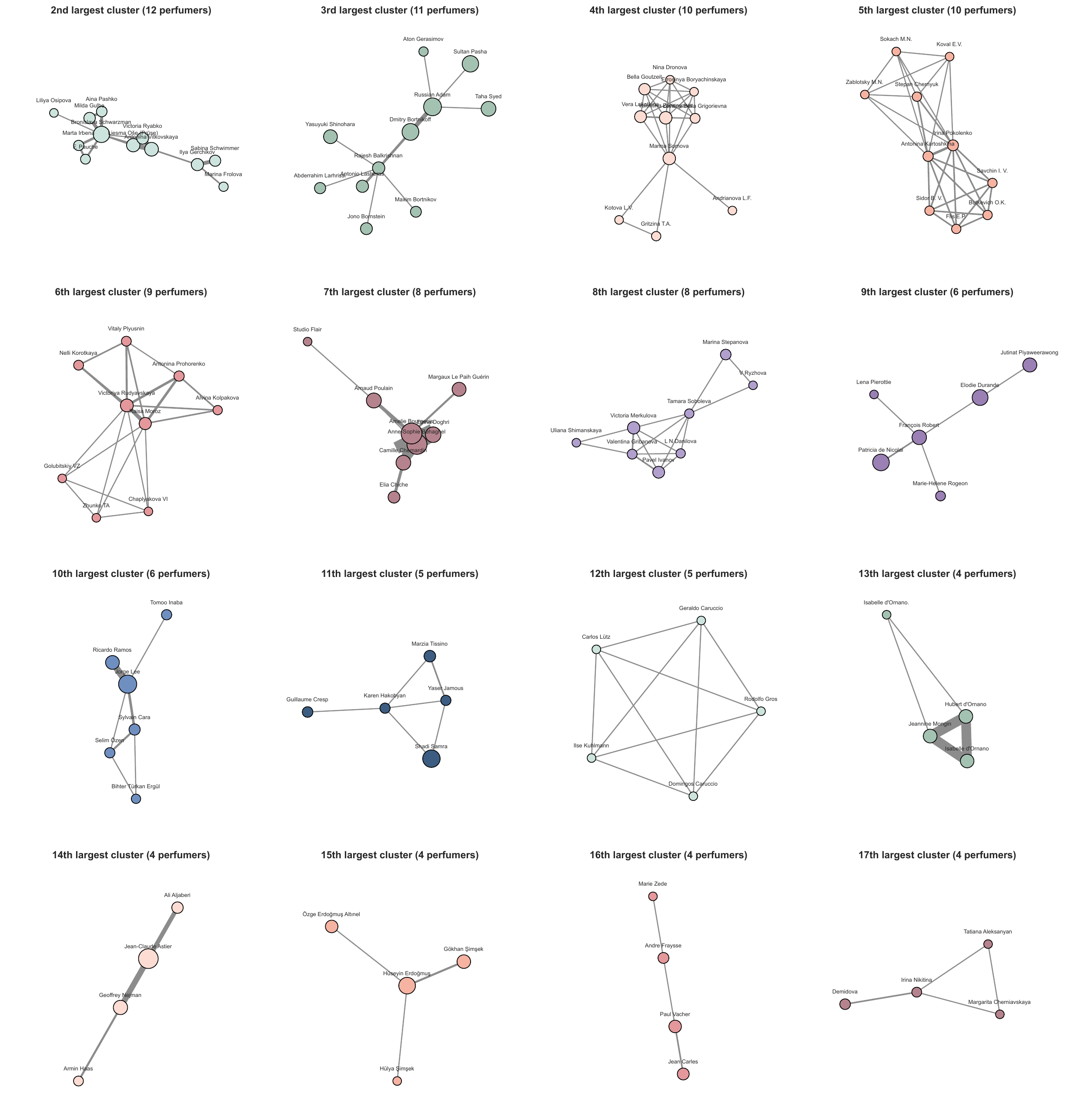}
\caption{\textbf{Smaller clusters of the perfumer collaboration network.} The 16 smaller connected components of perfumers collaboration network, the largest collaboration network is shown in Figure 4B. Node size is the number of perfumes made and node colour the collaboration community. Edges are black within a community and gray between, with thickness proportional to co-created perfumes. Labels indicate the leading perfumers.}\label{fig:S11}
\end{figure}

Using releases from 1900–2024, we kept countries with at least 30 perfumes and described each country by its share of the 40 most common notes (Figure~\ref{fig:S3}A) or accords (Figure~\ref{fig:S3}B), converted to a log location-quotient relative to the global average and z-scored across countries. We averaged these profiles within each of the six geographic regions and plotted them on overlaid radar plots. The ten axes per panel are the items with the highest PCA loadings (by combining absolute values of PC1 and PC2). The geographic share of variance is a one-way PERMANOVA pseudo-$R^2$ \cite{anderson2001new}. In sum, geographic region explains a modest but significant share of national scent profiles, 13\% for notes and 17\% for accords (Figure~\ref{fig:S3}).

\section{Gender coding of notes and accords} 
We constrain the data to two consecutive 25 year windows, an early one (1975–1999) and a recent one (2000–2024), and map each perfume's gender label to women, men, or unisex. For every note and accord we measure its gender balance among gendered perfumes by excluding unisex and using $\mathrm{(n_{women}- n_{men}) / (n_{women} + n_{men})}$. The balance is computed separately in each window for every note carried by at least 40 gendered perfumes in both windows. To test whether a note's coding changed, we apply a 2×2 chi-square (women vs men, early vs recent) and control the false-discovery rate across all tests with the Benjamini–Hochberg procedure. In Figure \ref{fig:S4} we plot every note and accord that experienced a significant changes, and show the notes and accords by star if they crossed gender balance line between the two windows.

\section{Semantic novelty}
As a complementary measure of novelty, we computed each perfume's novelty in embedding space. Each note was represented by its Word2Vec vector \cite{mikolov2013efficient}, and each perfume by the mean of its note vectors $\mathbf{v}$. We define semantic novelty by the cosine distance between a perfume's vector and the mean vector $\bar{\mathbf{v}}$ of all preceding perfumes,

\begin{equation}
\mathrm{Semantic~Novelty} = 1 - \frac{\mathbf{v}\cdot\bar{\mathbf{v}}}{\lVert\mathbf{v}\rVert\,\lVert\bar{\mathbf{v}}\rVert}
\end{equation}

\noindent perfumes whose composition is semantically distant from the past perfumes score higher in their semantic novelty. Scores were aggregated by year as the annual mean and are shown in \ref{fig:S5}C.

\section{Prestige effect on style adoption}

To ask which partner adopts the other's style when a higher and a lower prestige perfumer collaborate, we performed same calculation as done for Figs~4D-E. To measure this asymmetry in adoption, we split each ordered pair with history of direct collaborators to low and high prestige groups, and used pair's first co-creation years as the anchor year.

\begin{figure}[!htbp]
\centering
\includegraphics[width=0.7\textwidth]{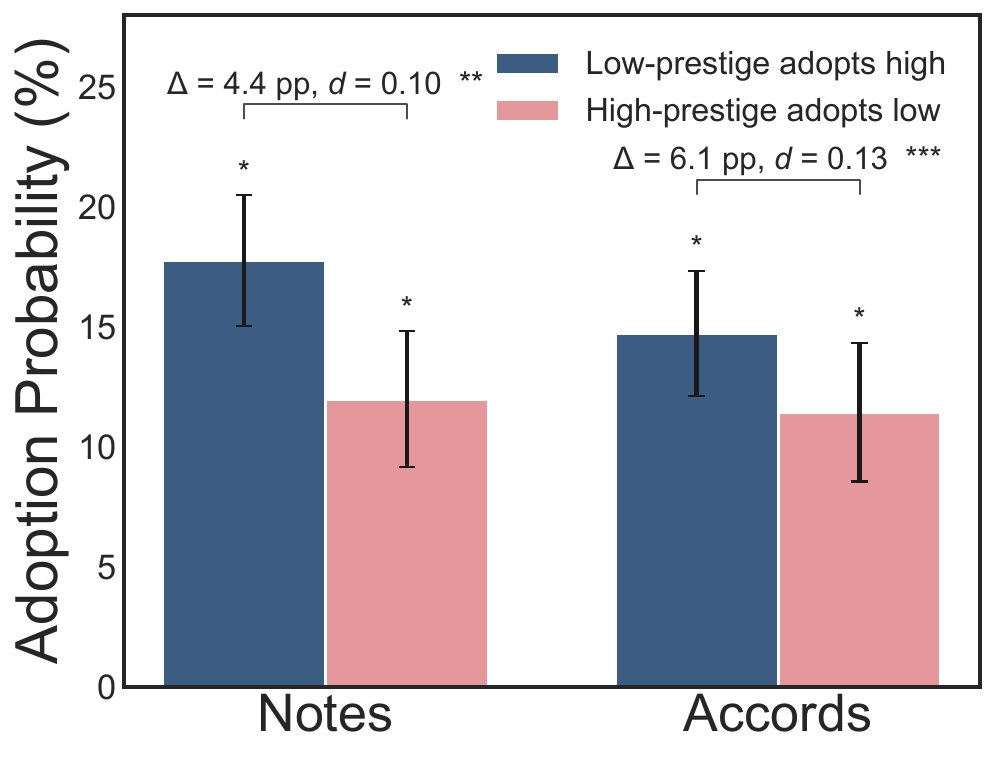}
\caption{\textbf{Prestige effect on style adoption between direct collaborators.} For accords and notes, the relative change in the probability that the lower-prestige partner adopts the higher-prestige partner's prior style versus the reverse. Prestige is each perfumer's eigenvector centrality measured before the collaboration year. Asterisks show bars differing from the null ($|z|\ge 1.96$). Brackets show the gap in adoption probability, Cohen's d, and significance.}\label{fig:S12}
\end{figure}

\begin{figure}[!htbp]
\centering
\includegraphics[width=1\textwidth]{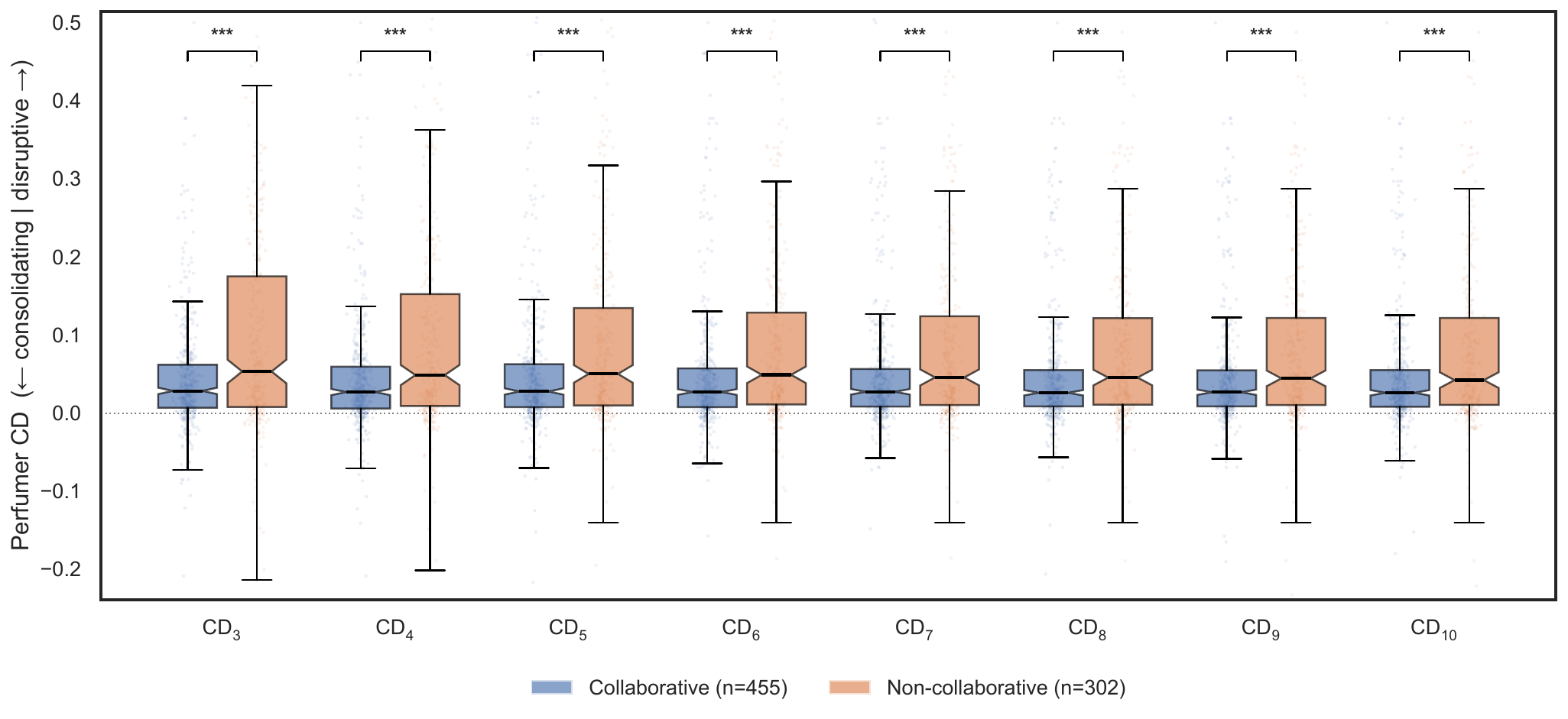}
\caption{\textbf{CD index for collaborative versus non-collaborative perfumers.} Per-perfumer mean consolidation-disruption (CD) index across the $CD_3-CD_{10}$ windows, split by whether a perfumer ever collaborated (blue) or always worked solo (orange), p-value is calculated with two-sided Mann-Whitney $U$.}\label{fig:S13}
\end{figure}

Here prestige is the time-sliced eigenvector centrality also known as prestige score. For a pair connecting in any given year, we recompute eigenvector centrality on the collaboration network as it existed before that year (edges with first-collaboration less than that year), so it reflects prestige at the moment of collaboration rather than eventual prestige. Within a pair, the partner with the higher pre-collaboration centrality is the high-prestige perfumer. A pair contributes an UP observation when the adopter is the lower-prestige partner and a DOWN observation when the adopter is the higher-prestige partner. Pairs in which both partners are newcomers with no prior collaborations are dropped.  For each direction we computed the adoption-probability relative change exactly as in Figs. 4D,E. The between-direction difference $\Delta = P_{\mathrm{obs}}^{\mathrm{UP}} - P_{\mathrm{obs}}^{\mathrm{DOWN}}$ was assessed with $z$-test. The relative change in adoption probability of accords and notes for each direction are shown in Figure~\ref{fig:S12}. We ran $n = 5000$ simulations and a direction differing from the null ($|z|\ge 1.96$) is shown by asterisks, the gap in adoption probability between directions, Cohen's d as the effect size, and its significance are shown by bracket (Figure~\ref{fig:S12}).

\section{IFRA restriction of oakmoss as a natural experiment}
The main text shows a population-level co-evolution, namely that over the twentieth century perfumes became simpler and more novel. This association between complexity and novelty is weak (Spearman $\rho$ of $-0.13$ for pairs and $-0.15$ for triplets, respectively), so the descriptive trade-off cannot on its own establish that constraint on the perfume formula causes exploration and a boost in novelty. To obtain a causal understanding we exploit an exogenous shock to the perfume ingredient. We investigate the example of the restriction of oakmoss by the International Fragrance Association (IFRA) \cite{marie2011regulatory}. Oakmoss is an important natural ingredient for the chypre olfactory family, and because it contains potent skin sensitizers, it has been among the most heavily regulated naturals in perfumery \cite{marie2011regulatory}. IFRA tightened its permitted level in a documented sequence of increasing severity. A first standard limited oakmoss absolute to 0.6\% of the finished product in 1988, this number reduced to 0.1\% in 1997, and 0.1\% limited on dehydroabietic acid in 2001, and a cap of 100 ppm implemented on atranol and chloroatranol in 2008 that effectively forced reformulation of oakmoss-built perfumes.

We classify each dated perfume by whether its note list contains oakmoss. Oakmoss perfumes form the treated group ($n = 2{,}040$) and all others the controls ($n = 12{,}923$). We focus in the window from 1970 to 2024. For each restriction year we fit a difference-in-differences model

\begin{equation}
y_i = \beta_0 + \beta_1\,\mathrm{treat}_i + \beta_2\,\mathrm{post}_i + \Delta\,(\mathrm{treat}_i \times \mathrm{post}_i) + \varepsilon_i,
\end{equation}

\noindent where $y$ is the outcome (i.e., complexity or novelty), and $\mathrm{treat}$ and $\mathrm{post}$ are binary indicators. Here, $\mathrm{treat}_i = 1$ if perfume $i$ contains oakmoss and $0$ otherwise; $\mathrm{post}_i = 1$ if perfume $i$ was released in restriction year or later and $0$ otherwise. Because both are binary, their product $\mathrm{treat}_i \times \mathrm{post}_i$ equals $1$ only for oakmoss perfumes released after the restriction and $0$ for every other perfume, so its coefficient $\Delta$ is the difference-in-differences estimate. $\Delta$ equals the before-to-after change in the treated group minus the corresponding change in the controls, and therefore removes both the constant level difference between the groups, as oakmoss perfumes are inherently more complex, and any era-wide trend common to both. A negative $\Delta$ for complexity and a positive $\Delta$ for novelty indicate that the restriction made oakmoss perfumes simpler and more novel relative to unaffected perfumes. The term $\varepsilon_i$ is a perfume-specific error capturing all variation in $y_i$ not explained by group membership and period. Identification of $\Delta$ as a causal effect requires only that this error has conditional mean zero, $\mathbb{E}[\varepsilon_i \mid \mathrm{treat}_i, \mathrm{post}_i] = 0$, which is the parallel-trends assumption. This does not require the error to have equal variance in the two groups. We estimate the model by ordinary least squares.

\begin{figure}[!htbp]
\centering
\includegraphics[width=1\textwidth]{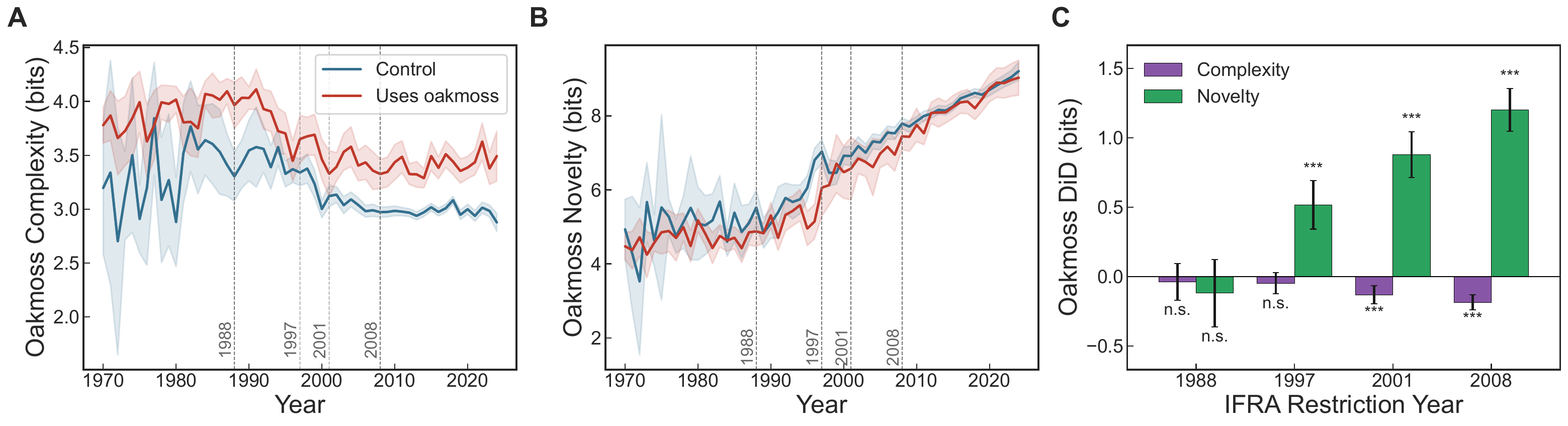}
\caption{\textbf{Exogenous regulatory restrictions effect on perfume complexity and novelty.} \textbf{(A)} Annual mean perfume complexity (Shannon entropy of the note distribution, in bits) and \textbf{(B)} annual mean perfume novelty (mean co-occurrence surprisal of note pairs and triples, in bits) for treated (red) and control (blue) perfumes, with 95\% confidence bands (treated n = 2{,}040, control n = 12{,}923). Dashed lines indicate the restriction years (Oakmoss limit of 0.6\% in 1988, 0.1\% in 1997, a dehydroabietic-acid limit in 2001, and an atranol/chloroatranol cap in the 2008 \cite{marie2011regulatory}). \textbf{(C)} Difference-in-differences for complexity and novelty at each oakmoss restriction year, with 95\% confidence intervals and significance  with $p$-values FDR-corrected by Benjamini-Hochberg.}\label{fig:S14}
\end{figure}

The estimates show a clear restriction induced response (Figure \ref{fig:S14}). The 1988 limit did not produce any detectable effect on either outcome. As the restriction tightened the effects grew monotonically in magnitude and became significant, reaching their maximum at the 2008, where oakmoss perfumes fell by 0.19 bits more in complexity and rose by 1.20 bits more in novelty than controls. All significant effects survive Benjamini-Hochberg correction for multiple comparisons (Figure \ref{fig:S14}).

\newpage

\newpage

\begin{table}[ht]
\centering
\small
\caption{Summary statistics of the three perfume datasets. The \textbf{Total} column reports unique entities across all three datasets after cross-dataset name matching; a dash (---) marks a metric not available in that dataset.}
\label{tab:dataset_comparison}
\begin{tabular}{lcccc}
\toprule
\textbf{Metric} & \textbf{Fragrantica} & \textbf{Aromo} & \textbf{Parfumo} & \textbf{Total} \\
\midrule
Perfumes & 92,589 & 78,409 & 59,280 & 169,759 \\
Brands & 4,780 & 5,936 & 1,450 & 8,332 \\
Perfumers & 2,475 & 548 & 1,258 & 3,050 \\
Unique canonical notes & 1,553 & 1,394 & 3,989 & 5,510 \\
Unique canonical accords & 91 & 23 & 21 & 111 \\
Year range & 1533--2025 & 1792--2024 & 1709--2024 & 1533--2025 \\
With release year & 76,632 & 31,619 & 38,984 & --- \\
With overall rating & 92,589 & 37,006 & 30,027 & --- \\
With main accords & 90,806 & 72,846 & 32,196 & --- \\
With notes / fragrance pyramid & 90,934 & 72,846 & 31,152 & --- \\
With perfumer & 39,700 & 6,320 & 20,530 & --- \\
Perfumes with names & 92,583 & 78,409 & 59,280 & --- \\
Gender categories & 3 & --- & --- & --- \\
Countries of origin & 99 & --- & --- & --- \\
Market segments & 5 & 8 & --- & --- \\
Concentration types & --- & 8 & 411 & --- \\
Mean rating (1--10 scale) & 7.92 & 7.84 & 7.35 & --- \\
Overall rating votes & 16,441,597 & 221,020 & 1,821,995 & --- \\
Note votes & 25,871,640 & --- & --- & --- \\
Sillage votes & 8,243,094 & --- & --- & --- \\
Longevity votes & 7,353,488 & --- & --- & --- \\
Price-value votes & 4,560,926 & --- & --- & --- \\
Gender votes & 5,147,769 & --- & --- & --- \\
Collection signals (had/have/want) & 46,377,488 & --- & --- & --- \\
Perfumes with user-submitted notes & --- & 1,436 & --- & --- \\
Perfumes with resemblance annotations & 69,549 & --- & --- & --- \\
Total resemblance annotations & 545,732 & --- & --- & --- \\
\bottomrule
\end{tabular}
\end{table}  
%%%%%%%%%%%%%%%% START OF SUPPLEMENT %%%%%%%%%%%%%%%
\newpage

\end{document}